\documentclass[journal]{IEEEtran}
\usepackage{cite}
\usepackage{amsmath,amssymb,amsfonts}
\usepackage{graphicx}
\usepackage{textcomp}
\usepackage{xcolor}
\usepackage{multirow}
\usepackage{booktabs}
\usepackage{hyperref}
\usepackage{tikz}
\usepackage{comment}
\usepackage{algorithm}% http://ctan.org/pkg/algorithm
\usepackage{algpseudocode}% http://ctan.org/pkg/algorithmicx
\usepackage{authblk}
\usepackage{subcaption}
\usepackage{threeparttable}
\usepackage[font=small,labelfont=bf]{caption}
\definecolor{ao(english)}{rgb}{0.0, 0.5, 0.0}

\newcommand{\blue}[1]{\textcolor{blue}{#1}} 
\hypersetup{
    colorlinks=true,
    linkcolor=blue,
    filecolor=magenta,      
    urlcolor=cyan,
    citecolor = magenta,
    pdfpagemode=FullScreen,
}
\usepackage{color}

\DeclareMathOperator*{\Motimes}{\text{\raisebox{0.25ex}{\scalebox{0.8}{$\bigotimes$}}}}

\ifCLASSINFOpdf
\else
\fi
\begin{document}
%\title{Demystifying Neuro-Symbolic AI with Workload Characterization and Software-Hardware Co-Design}
%\title{Towards Efficient Neuro-Symbolic AI: Workload Characterization and Hardware Acceleration}
\title{\fontsize{23.6}{28.0}\selectfont{Cross-Layer Design for Neuro-Symbolic AI: From Workload Characterization to Hardware Acceleration}}

\title{Towards Efficient Neuro-Symbolic AI: From Workload Characterization to Hardware Architecture}

\author[$1$]{Zishen~Wan}
\author[$1$]{Che-Kai Liu}
\author[$1$]{Hanchen Yang}
\author[$1$]{Ritik Raj}
\author[$1$]{Chaojian Li}
\author[$1$]{Haoran You}
\author[$1$]{Yonggan Fu}
\author[$1$]{\\Cheng Wan}
\author[$1$]{Sixu Li}
\author[$2$]{Youbin Kim}
\author[$3$]{Ananda Samajdar}
\author[$1$]{Yingyan (Celine) Lin}
\author[$1,2$]{Mohamed Ibrahim}
\author[$2$]{\\Jan M. Rabaey}
\author[$1$]{Tushar Krishna} 
\author[$1$]{Arijit Raychowdhury\vspace{-0.05in}} 

\affil[$ $]{\normalsize \textit{$^{1}$Georgia Institute of Technology\hspace{0.1in} $^{2}$University of California, Berkeley\hspace{0.1in} $^{3}$IBM Research}\vspace{-0.3in}}

\markboth{IEEE Transactions on Circuits and Systems for Artificial Intelligence,~Vol.~xx, No.~xx, September~2024}%
{Shell: Running title}
% {Shell \MakeLowercase{\textit{et al.}}: Bare Demo of IEEEtran.cls for IEEE Journals}



\maketitle

\begin{abstract}
The remarkable advancements in artificial intelligence (AI), primarily driven by deep neural networks, are facing challenges surrounding unsustainable computational trajectories, limited robustness, and a lack of explainability. To develop next-generation cognitive AI systems, neuro-symbolic AI emerges as a promising paradigm, fusing neural and symbolic approaches to enhance interpretability, robustness, and trustworthiness, while facilitating learning from much less data. Recent neuro-symbolic systems have demonstrated great potential in collaborative human-AI scenarios with reasoning and cognitive capabilities.
In this paper, we aim to understand the workload characteristics and potential architectures for neuro-symbolic AI. We first systematically categorize neuro-symbolic AI algorithms, and then experimentally evaluate and analyze them in terms of runtime, memory, computational operators, sparsity, and system characteristics on CPUs, GPUs, and edge SoCs. Our studies reveal that neuro-symbolic models suffer from inefficiencies on off-the-shelf hardware, due to the memory-bound nature of vector-symbolic and logical operations, complex flow control, data dependencies, sparsity variations, and limited scalability. Based on profiling insights, we suggest cross-layer optimization solutions and present a hardware acceleration case study for vector-symbolic architecture to improve the performance, efficiency, and scalability of neuro-symbolic computing. Finally, we discuss the challenges and potential future directions of neuro-symbolic AI from both system and architectural perspectives.
\end{abstract}


% \fi
\IEEEpeerreviewmaketitle

\section{Introduction}
% \vspace{-0.3em}
\label{sec:intro}
The remarkable advancements in AI have had a profound impact on our society. These advancements are primarily driven by deep neural networks and a virtuous cycle involving large networks, extensive datasets, and augmented computing power. As we reap the benefits of this success, there is growing evidence that continuing our current trajectory may not be viable for realizing AI's full potential. 
% Three primary concerns are worth noting. 
First, the escalating computational requirements and energy consumption associated with AI are on an unsustainable trajectory~\cite{wu2022sustainable}, threatening to reach a level that could stifle innovation by restricting it to fewer organizations. Second, the lack of robustness and explainability remains a significant challenge, likely due to inherent limitations in current learning methodologies~\cite{wan2021analyzing,debenedetti2023scaling}. Third, contemporary AI systems often operate in isolation with limited collaboration among humans and other AI agents. Hence, it is imperative to develop next-generation AI paradigms that address the growing demand for enhanced efficiency, explainability, and trust in AI systems.

\begin{figure}[t!]
    \includegraphics[width=.95\columnwidth]{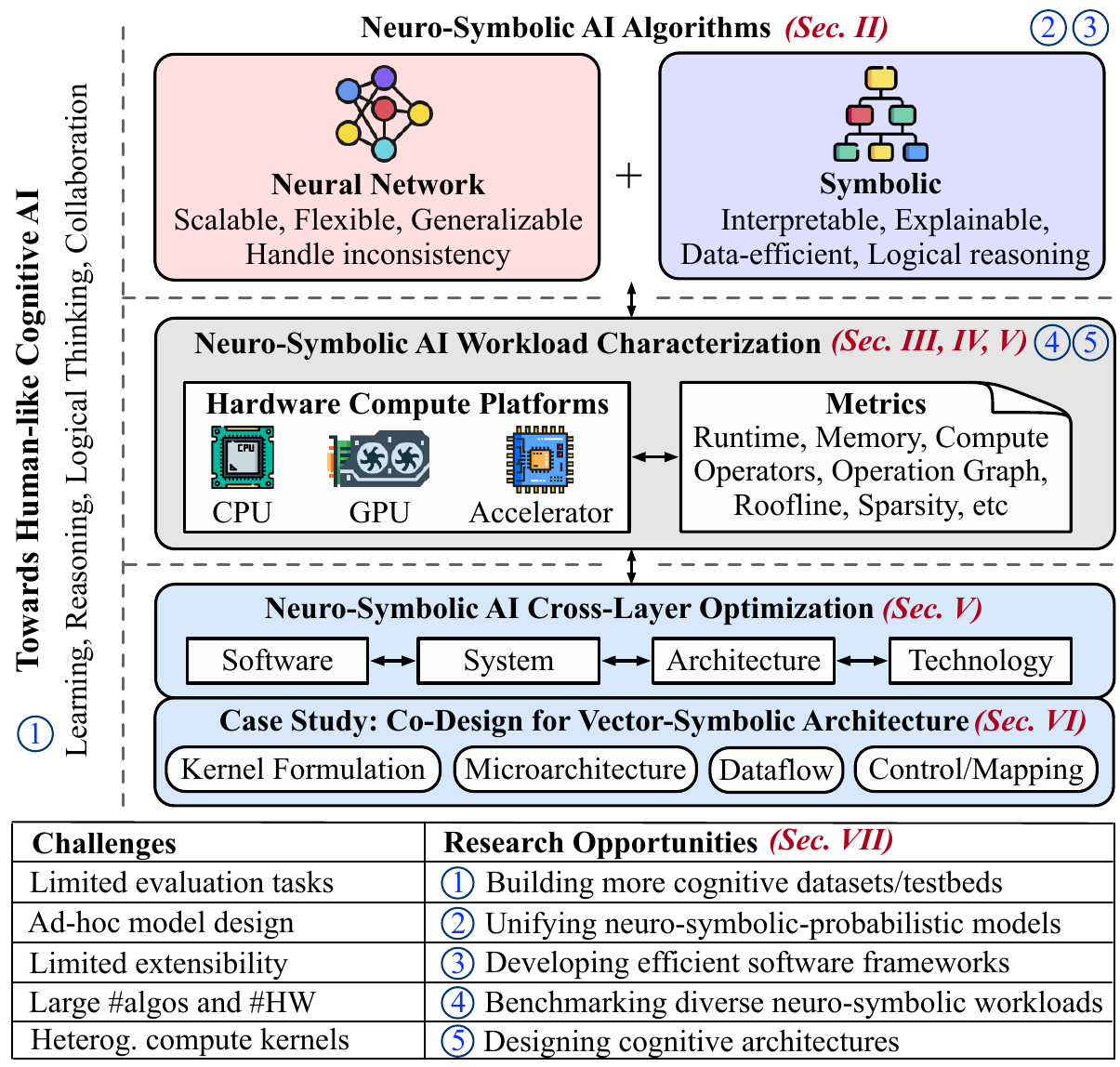}
        % \vspace{-1em}
        \centering
         \caption{Overview of neuro-symbolic AI systems, workload characterizations, optimization solutions, challenges, and research opportunities in improving the performance of next-generation cognitive AI.}
        \label{fig:overview}
        \vspace{-10pt}
\end{figure}

% what is NS-AI, its advatanges
% Neuro-symbolic AI (NSAI) is an emerging AI paradigm that combines neural, symbolic, and probabilistic approaches to AI to improve explainability and robustness and enable learning from less data (Fig.~\ref{fig:overview}). Neural methods demonstrate remarkable effectiveness in extracting intricate features from data for tasks like natural language processing and object detection. Meanwhile, symbolic methods bring explainability and reduce the need for training data by incorporating known models of the physical world, and probabilistic methods allow cognitive systems to better deal with uncertainty and therefore be more robust to unstructured conditions. A synergistic integration of neural, symbolic, and probabilistic methods makes NSAI a promising paradigm that unlocks the third wave of AI~\cite{garcez2023neurosymbolic}.
Neuro-symbolic AI~\cite{wan2024towards_2} represents an emerging AI paradigm that integrates the neural and symbolic approaches with probabilistic representations to enhance explainability, robustness and facilitates learning from much less data in AI (Fig.~\ref{fig:overview}). Neural methods are highly effective in extracting complex features from data for vision and language tasks. On the other hand, symbolic methods enhance explainability and reduce the dependence on extensive training data by incorporating established models of the physical world, and probabilistic representations enable cognitive systems to more effectively handle uncertainty, resulting in improved robustness under unstructured conditions. The synergistic fusion of neural and symbolic methods positions neuro-symbolic AI as a promising paradigm capable of ushering in the third wave of AI~\cite{garcez2023neurosymbolic,wan2024towards}.

% conenct to application
% NSAI unlocks many possibilities of scenarios that acquire human-like communication and reasoning capabilities and be able to recognize, classify and adapt to new situations autonomously. Besides its superior performance compared to ML models in tasks such as image and video question answering~\cite{maoneuro,yang2020neurasp}, NSAI has great potential in advancing real-time responses, energy efficiency, explainability and trustworthiness of collaborative human-AI applications (e.g., collaborative robotics, mixed-reality systems and human-AI interaction in the metaverse), where robots can seamlessly interact with humans in complex environments, AI agents can reason and make decisions in a robust and explainable manner, and intelligence embedded pervasively and untethered from the cloud.
Neuro-symbolic AI promises possibilities for systems that acquire human-like communication and reasoning capabilities, enabling them to recognize, classify, and adapt to new situations autonomously. For example, neuro-vector-symbolic architecture~\cite{hersche2023neuro} is able to reach 98.8\% accuracy on spatial-temporal reasoning tasks, greatly surpassing human performance (84.4\%), neuro-only ResNet (53.4\%) and GPT-4 performance (89.0\%). 
In addition to its superior performance in vision and language~\cite{maoneuro}, neuro-symbolic AI holds significant potential for enhancing explainability and trustworthiness of collaborative human-AI applications~\cite{gu2023conceptgraphs}. These applications include collaborative robotics, mixed-reality systems, and human-AI interactions, where robots can seamlessly interact with humans in environments, agents can reason and make decisions in an explainable manner, and intelligence is pervasively embedded and untethered from the cloud. 

Despite the promising algorithmic performance, the higher memory intensity, greater kernel heterogeneity, and access pattern irregularity of neuro-symbolic computing lead to an increasing divergence from the current hardware roadmap that largely optimizes for matrix multiplication and convolution~\cite{samajdar2020systematic,kwon2021heterogeneous,wu2023highlight,ramachandran2024algorithm,fan2024benchmarking} and leads to severe inefficiencies and underutilization of hardware. Therefore, understanding its computational and memory demands is essential for efficient processing on both general-purpose and custom hardware.

% this paper
% In this paper, we systematically survey, evaluate, and analyze NSAI systems - a next-generation AI paradigm.
% First, we review the state-of-the-art NSAI systems and categorize them in a structured perspective (Sec.~\ref{sec:survey}).
% Second, we analyze various NSAI workloads on hardware platforms and analyze their runtime characteristics and underlying compute operators (Sec.~\ref{sec:profile}).
% Lastly, we discuss challenges and opportunities for NSAI system research, and our view of the road ahead (Sec.~\ref{sec:challenge}).
% This is the \emph{first} paper that assesses NSAI from a system and architecture perspective, aiming to inspire the design of next-generation cognitive computing systems through synergistic advances in NSAI algorithms, systems, architecture, and algorithm-hardware co-design.
Our goal in this work is to quantify the workload characteristics and potential system architecture for neuro-symbolic AI. Built on our work~\cite{wan2024towards_2,ibrahim2024efficient}, we first conduct a systematic review and categorize state-of-the-art neuro-symbolic AI workloads in a structured manner (Sec.~\ref{sec:survey}). We then characterize seven representative neuro-symbolic workloads on general-purpose and edge platforms, analyzing their runtime, memory, compute operators, operation graph, hardware utilization, and sparsity characteristics (Secs.~\ref{sec:model}, \ref{sec:method}, \ref{sec:results}). Our workload characterization reveals several key observations and insights, including the following:

    % \item We systematically review and categorize state-of-the-art neuro-symbolic AI workloads in a structured manner (Sec.~\ref{sec:survey}).
    % \item We characterize seven representative neuro-symbolic workloads on general-purpose and edge platforms, and analyze their runtime, memory, operation graph, sparsity characteristics, and underlying neural and symbolic operators. Our analysis provides new insights: 

\begin{itemize}
    \item Neuro-symbolic AI models typically exhibit high latency compared to neural models, prohibiting them from real-time applications.
    \item The neural components mainly consist of MatMul and Convs, while the symbolic components are dominated by vector/element-wise and logical operations. The low ALU utilization, low cache hit rates, and high volume of data movement of symbolic operations make them inefficient on CPUs/GPUs and may result in system bottlenecks.
    % \item The hardware inefficiency of symbolic operations typically is due to low ALU utilization, low cache hit rates, and high volume of data movement.
    \item The neural workloads are compute-bounded while the symbolic workloads are typically memory-bounded and face potential scalability issues.
    \item The symbolic operations may depend on neural results or need to compile into the neural structure, thus lying on the critical path of end-to-end neuro-symbolic systems.
    \item Some neural and vector-symbolic components demonstrate a high level of unstructured sparsity with variations under different task scenarios and attributes.
\end{itemize}

Inspired by our workload profiling insights, we recommend several cross-layer software and hardware optimization solutions to improve the efficiency and scalability of neuro-symbolic systems (Sec.~\ref{sec:results}). Specifically, we leverage vector-symbolic architecture as a case study and present a hardware acceleration methodology, including kernel formulation, microarchitecture, dataflow, and control schemes (Sec.~\ref{sec:hardware}).
Finally, we explore the research opportunities in neuro-symbolic computing and share our outlook on the road ahead (Sec.~\ref{sec:challenge}).

% \textbullet~We systematically review and categorize state-of-the-art neuro-symbolic AI workloads in a structured manner (Sec.~\ref{sec:survey}).

% \textbullet~We characterize seven representative neuro-symbolic workloads on general-purpose and edge platforms, analyzing their runtime, memory, operation graph, sparsity characteristics, and underlying neuro and symbolic operators. Our analysis provides new insights: (1) Neuro-symbolic computing suffers from inefficient processing on off-the-shelf hardware due to complex flow control, holographic vector operation, and data transfer overhead; (2) Neuro workloads are compute-bound while symbolic workloads are memory-bound and face potential scalability issues; (3) Symbolic operations typically rely on the neural results or need integration into the neuro structure, thus lying on the critical path; (4) Neuro-symbolic workloads demonstrate a high level of unstructured sparsity with variations under different scenarios (Secs.~\ref{sec:model}, \ref{sec:method}, \ref{sec:results}).

% \textbullet~Inspired by our workload profiling insights, we propose cross-layer software and hardware optimization solutions to improve the efficiency and scalability of neuro-symbolic systems, including model compression, sparsity optimization, multi-level parallelism, heterogeneous and specialized computing elements, and a memory-centric design (Sec.~\ref{sec:optimization}).

% \textbullet~We explore the research opportunities in neuro-symbolic computing and share our outlook on the road ahead (Sec.~\ref{sec:challenge}).

To the best of our knowledge, this is one of the \emph{first} works to characterize neuro-symbolic computing from both system and architectural perspectives, and enable its efficient and scalable execution. We aim to inspire the design of next-generation cognitive computing systems through synergistic advancements in neuro-symbolic algorithms, systems, architecture, and algorithm-hardware co-design.

 % \emph{However, neuro-symbolic systems exhibit divergent compute and memory features than DNNs and make current deep learning architecture inefficient, this paper thus takes the first step to characterize neuro-symbolic workloads to enable their efficient and scalable execution.}

 % \emph{Neuro-symbolic AI shows the promising potential to be integrated into these systems and enable trustworthy applications. It is thus highly desirable to understand and optimize its system characteristics and performance.}
% \vspace{-0.5em}
\section{Neuro-Symbolic AI Algorithms}
% \vspace{-0.3em}
\label{sec:survey}

\begin{figure*}[t!]
\centering
\begin{minipage}[b]{\linewidth}
    \centering
    \includegraphics[width=.7\columnwidth]{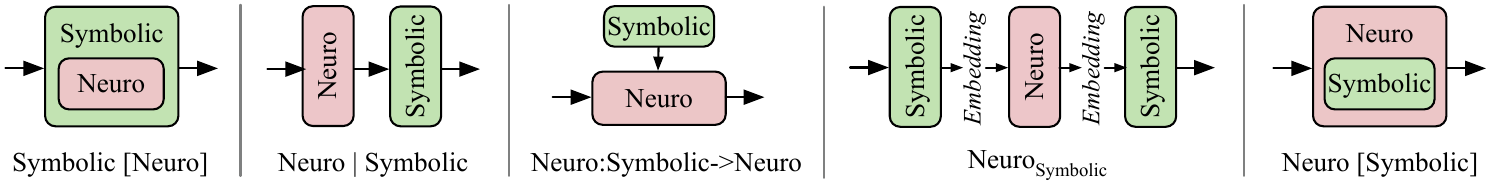}
\end{minipage}% 
 \vspace{2pt}
\begin{minipage}[b]{\linewidth}
\centering
\huge
    \renewcommand*{\arraystretch}{1.1}
\resizebox{.93\linewidth}{!}{%
% \begin{table*}[t!]
% \vspace{-5pt}
\centering
% \caption{xxx}
% \renewcommand*{\arraystretch}{1.05}
\setlength\tabcolsep{5pt}
% \resizebox{\linewidth}{!}{%
\begin{tabular}{c|c|c|c|c}
\hline
\textbf{\begin{tabular}[c]{@{}c@{}}Category\end{tabular}} & \textbf{\begin{tabular}[c]{@{}c@{}}Category Description\end{tabular}}  & \multicolumn{1}{c|}{\textbf{Neuro-Symbolic Algorithm}}                             & \textbf{\begin{tabular}[c]{@{}c@{}} Underlying Operation\end{tabular}} & \multicolumn{1}{c}{\textbf{If Vector}}                \\ \hline
\textbf{Symbolic{[}Neuro{]}}  & End-to-end \textbf{symbolic} system that uses \textbf{neural} models internally as a subroutine                                   & \textbf{AlphaGo} \cite{silver2017mastering}     &    NN, MCTS                                                                                  & Vector                 \\ \hline
\multirow{8}{*}{\textbf{Neuro$|$Symbolic}} & \multirow{8}{*}{\begin{tabular}[c]{@{}c@{}}Pipelined system that integrates \textbf{neural} and \textbf{symbolic} components where \\each component specializes in complementary tasks within the whole system\end{tabular}} & \textbf{NVSA}~\cite{hersche2023neuro}  & NN, mul, add, circular conv.       &   Vector \\ \cline{3-5} 
               &      & \textbf{NeuPSL} \cite{pryor2022neupsl}   &  NN, fuzzy logic      &              Vector  \\ \cline{3-5} 
              &  & \textbf{NSCL}  \cite{maoneuro} & NN, add, mul, div, log    &      Vector \\ \cline{3-5} 
              &       & \textbf{NeurASP} \cite{yang2020neurasp} &  NN, logic rules     &   Non-Vector \\ \cline{3-5} 
              &       & \textbf{ABL}  \cite{dai2019bridging} &NN, logic rules    &       Non-Vector \\ \cline{3-5} 
               &      & \textbf{NSVQA}  \cite{yi2018neural}  &  NN, pre-defined objects      &     Non-Vector \\ \cline{3-5}
              &       & \textbf{VSAIT}  \cite{theiss2022unpaired}  &  NN, binding/unbinding    & Vector \\ \cline{3-5} 
              &      & \textbf{PrAE}  \cite{zhang2021abstract}  &  NN, logic rules, prob. abduction      &  Vector \\ \hline

\multirow{3}{*}{\begin{tabular}[l]{@{}l@{}}\textbf{Neuro:Symbolic}\textbf{$\rightarrow$Neuro}\end{tabular}}& \multirow{3}{*}{\begin{tabular}[l]{@{}l@{}}End-to-end \textbf{neural} system that compiles \textbf{symbolic} knowledge externally\end{tabular}} & \textbf{LNN} \cite{riegel2020logical}  &  NN, fuzzy logic               & Vector\\ \cline{3-5} 

              &       & \begin{tabular}[l]{@{}l@{}}\textbf{Symbolic Math}~\cite{lampledeep}\end{tabular} &  NN                                                                            & Vector \\ \cline{3-5} 
                   
              &       & \begin{tabular}[l]{@{}l@{}}\textbf{Differentiable ILP}~\cite{evans2018learning}\end{tabular} &  NN, fuzzy logic                                                                                     & Vector \\ \hline
\multirow{2}{*}{\textbf{$\mbox{Neuro}_{\mbox{Symbolic}}$ }} & \multirow{2}{*}{\begin{tabular}[c]{@{}c@{}}Pipelined system that maps \textbf{symbolic} first-order logic onto embeddings \\serving as soft constraints or regularizers for \textbf{neural} model\end{tabular}} & \textbf{LTN} \cite{badreddine2022logic}  & NN, fuzzy logic                                                                    & Vector \\ \cline{3-5} 
               &      & \begin{tabular}[l]{@{}l@{}}\textbf{DON}~\cite{hohenecker2020ontology}\end{tabular}  & NN  & Vector  \\ \hline
\multirow{3}{*}{\textbf{Neuro{[}Symbolic{]}} }  &\multirow{3}{*}{End-to-end \textbf{neural} system that uses \textbf{symbolic} models internally as a subroutine}                 & \begin{tabular}[l]{@{}l@{}}\textbf{GNN+attention}~\cite{lamb2020graph}\end{tabular} & NN, SpMM, SDDMM & Vector\\ \cline{3-5} 
&  & \textbf{ZeroC} \cite{wu2022zeroc}  & NN (energy-based model, graph)& Vector \\  \cline{3-5} 
&  & \textbf{NLM} \cite{dongneural}  &  NN, permutation         &  Vector \\  \hline
                  % & \textbf{SATNet}  \cite{wang2019satnet} &  & Yes \\ \hline
\end{tabular}}
\captionof{table}{Review of recent neuro-symbolic AI algorithms into five categories, with their underlying operations and vector formats.}
%\TK{in this table can we add two roes, one saying Neural and other saying Symbolic, and describe what the specific DNN and symbolic method in each is? Sec II-B text says some of this, trying to get it into table.} \ZW{updated the table, pls take a look}}
\vspace{-10pt}
\label{tab:summary}
\end{minipage}
\end{figure*}

% \caption{NS-AI taxonomy, operation, and scalability.}

In this section, we systematically review and categorize the recent research progress in neuro-symbolic AI algorithms.
\textbf{Overview.} Neuro-symbolic AI represents an interdisciplinary approach that synergistically combines symbolic reasoning with neural network (NN) learning to create intelligent systems, leveraging the complementary strengths of both to enhance the accuracy and interpretability of the resulting models. Given that neuro-symbolic algorithms incorporate symbolic and neural components, various paradigms can be categorized based on how these components are integrated into a cohesive system. Inspired by Henry Kautz's taxonomy~\cite{henry2020taxonomy}, we systematically categorize these algorithms into five paradigms (Tab.~\ref{tab:summary}). We elaborate on each of these paradigms below. Additionally, Tab.~\ref{tab:operation_example} provides examples of several underlying operations based on the categorization in Tab.~\ref{tab:summary}.%, which will be elaborated as follows.} 
% \textbf{Symbolic[Neuro]} refers to a comprehensive symbolic problem solver with loosly-coupled neural pattern recognition subroutines. Examples include DeepMind's AlphaGo \cite{silver2017mastering} and AlphaZero~\cite{zhang2020alphazero}, where the Monte-Carlo Tree Search is used as the symbolic problem solver with neural network-based state estimators. 
% such as policy and value networks.

% \textbf{Symbolic[Neuro]} refers to an intelligent system that empowers symbolic reasoning with the statistical learning %pattern recognition 
% capabilities of neural networks. These systems typically consist of a comprehensive symbolic problem solver that includes loosely-coupled neural subroutines for statistical learning. %recognizing patterns. 
% Examples include DeepMind's AlphaGo \cite{silver2017mastering} and AlphaZero~\cite{zhang2020alphazero}, which use Monte-Carlo Tree Search (MCTS) as the symbolic solver and neural network state estimators for learning statistical patterns. %pattern recognition.}
\textbf{Symbolic[Neuro]} refers to an intelligent system that empowers symbolic reasoning with the statistical learning 
capabilities of NNs. These systems typically consist of a comprehensive symbolic problem solver that includes loosely-coupled neural subroutines for statistical learning. 
Examples include DeepMind's AlphaGo \cite{silver2017mastering} and AlphaZero~\cite{zhang2020alphazero}, which use Monte-Carlo Tree Search (MCTS) as the symbolic solver and NN state estimators for learning statistical patterns. 

\textbf{Neuro$|$Symbolic} refers to a hybrid system that combines neural and symbolic components in a pipeline, where each component typically specializes in complementary tasks. To the best of our knowledge, the majority of neuro-symbolic algorithms fall into this category.
For example, IBM's neuro-vector-symbolic architecture (NVSA)~\cite{hersche2023neuro} uses an NN as the perception frontend for semantic parsing and a symbolic reasoner as the backend for probabilistic abductive reasoning on the RAVEN~\cite{zhang2019raven} and I-RAVEN~\cite{hu2021stratified} datasets. Probabilistic abduction and execution (PrAE) learner~\cite{zhang2021abstract} adopts a similar approach where the difference lies in features are first projected to high-dimensional vectors in NVSA, whereas PrAE utilizes the original features directly as the NN's input. 
% Vector symbolic architecture-based image-to-image translation (VSAIT)~\cite{theiss2022unpaired} first extract features with NN and use locality sensitive hashing (LSH) to encode images into high-dimensional space and finally performs symbolic operations. 
Other examples include vector symbolic architecture-based image-to-image translation (VSAIT)~\cite{theiss2022unpaired}, neuro-probabilistic soft logic (NeuPSL) \cite{pryor2022neupsl}, neural probabilistic logic programming (DeepProbLog)~\cite{manhaeve2021neural}, neuro-answer set programming (NeurASP)~\cite{yang2020neurasp}, neural symbolic dynamic reasoning~\cite{yi2020clevrer}, neural symbolic concept learner (NSCL) \cite{maoneuro}, abductive learning (ABL) \cite{dai2019bridging}, and neuro-symbolic visual question answering (NSVQA) \cite{yi2018neural} on the CLEVRER dataset~\cite{yi2020clevrer}.

\textbf{Neuro:Symbolic$\rightarrow$Neuro} approach incorporates symbolic rules into NNs to guide the learning process, where symbolic knowledge is compiled into the structure of neural models for enhancing the model interpretability. 
For instance, logical NNs (LNNs)~\cite{riegel2020logical} encode knowledge or domain expertise as symbolic rules (first-order logic or fuzzy logic) that act as constraints on the NN output. Other examples include the application of deep learning for symbolic mathematics \cite{lampledeep} and differentiable inductive logic programming (ILP) \cite{evans2018learning}.

\textbf{$\mbox{Neuro}_{\mbox{Symbolic}}$} is a type of hybrid approach that combines symbolic logic rules with NNs. It involves mapping symbolic logic rules onto embeddings that serve as soft constraints or regularizers on the NN's loss function.
Logical tensor networks (LTNs) \cite{badreddine2022logic}, for instance, use logical formulas to define constraints on the tensor representations, which have proven successful in knowledge graph completion tasks. These tasks aim to predict missing facts or relationships between entities. Other examples of this approach include deep ontology networks (DONs)~\cite{hohenecker2020ontology} and tensorization methods~\cite{garcez2019neural}. As inference is still governed by NNs, it remains a research question whether this approach will compromise interpretability.

\begin{table}[t!]
% \vspace{-0.5em}
\centering
\caption{Enumeration of the underlying operations based on Tab.~\ref{tab:summary}\vspace{3pt}.}
\renewcommand*{\arraystretch}{1.1}
\setlength\tabcolsep{2.2pt}
\resizebox{1\columnwidth}{!}{
\begin{tabular}{c|c}
\hline
\textbf{Underlying Operations}                   & \textbf{Examples}  \\ \hline
\multirow{2}{*}{\begin{tabular}[c]{@{}c@{}}Fuzzy logic\\ (LTN)\end{tabular}} & $F = \forall x (isCarnivor(s)) \rightarrow (isMammal(x))$  \\ 
                    &  $\{isCarnivor(s)$:$[0,1]$, $isMammal(x)$:$[1,0]$\} $\rightarrow F = [1,0]$ \\ \hline
Mul, Add, and Circular Conv.  & \multirow{2}{*}{$X_i \in \{+1, -1\}^d \rightarrow (X_i \cdot X_j) / (X_i + X_j)$} \\
(NVSA) & \\\hline
\multirow{3}{*}{\begin{tabular}[c]{@{}c@{}}Logic rules\\ (ABL)\end{tabular}} & Domain: $animal (dog). carnivore(dog).mammal(dog)$ \\ 
                    & Logical formula: $mammal(x)\wedge carnivore(x)$ \\
                    & ABL: $hypos(x): -animal(x),mammal(x),carnivore(x)$
                    \\ \hline
\multirow{2}{*}{\begin{tabular}[c]{@{}c@{}}Pre-defined objects\\ (NSVQA)\end{tabular}} & \texttt{equal\_color:} $(entry, entry) \rightarrow Boolean $ \\ 
                    & \texttt{equal\_integer:} $(number, number) \rightarrow Boolean $ \\ \hline
\end{tabular}}
\label{tab:operation_example}
\vspace{-10pt}
\end{table}

% \textbf{Neuro[Symbolic]} is the transpose of Symbolic[Neuro], in which the overall neural model incorporates symbolic reasoning by paying attention to specific symbolics at certain conditions. For instance, graph neural networks (GNNs) can possibly be a strong tool for representing symbolic expressions, while being endowed with attention mechanisms \cite{lamb2020graph}.

% \textbf{Neuro[Symbolic]} refers to a system that empowers neural networks with the explainability and robustness of symbolic reasoning. Unlike $\textbf{Symbolic[Neuro]}$, where symbolic reasoning is used to guide the neural model learning process, in $\textbf{Neuro[Symbolic]}$, the overall neural model incorporates symbolic reasoning by paying attention to specific symbolics at certain conditions. For instance, graph neural networks (GNNs) are often adopted as strong candidates for representing symbolic expressions when endowed with attention mechanisms \cite{lamb2020graph}. In particular, this attention mechanism can be used to incorporate symbolic rules into GNN models, enabling selective attention to pertinent %by selectively attending to relevant 
% symbolic information in the graph. Other examples include neural logic machines (NLM)~\cite{dongneural}.
\textbf{Neuro[Symbolic]} refers to a system that empowers NNs with the explainability and robustness of symbolic reasoning. Unlike $\textbf{Symbolic[Neuro]}$, where symbolic reasoning is used to guide the neural model learning process, in $\textbf{Neuro[Symbolic]}$, the neural model incorporates symbolic reasoning by paying attention to a specific symbolic at certain conditions. For instance, graph neural networks (GNNs) are adopted for representing symbolic expressions when endowed with attention mechanisms \cite{lamb2020graph}. In particular, this attention mechanism can be leveraged to incorporate symbolic rules into GNN models, enabling selective attention to pertinent  
symbolic information in the graph. 
% At inference, ZeroC first parses each image into respective concept graphs, preserving their explainability. 
Other examples include neural logic machines (NLM)~\cite{dongneural} and Zero-shot concept recognition and acquisition (ZeroC)~\cite{wu2022zeroc}. ZeroC leverages the graph representation where the constituent concept models are represented as nodes and their relations are represented by edges. 

Each neuro-symbolic category reflects different kernel operators and data dependencies. \emph{Therefore, this paper takes one of the first steps towards understanding its computing characteristics and aims to serve as a cornerstone for the design and deployment of future neuro-symbolic systems.}

% \textcolor{blue}{ToDo: add Vector Symbolic Architecture-based Image-to-Image Translation (VSAIT)~\cite{theiss2022unpaired}}

% \input{Tabs/tab_also_survey}

% \subsection{LNN}
% \label{subsec:lnn}
% The performance will be bad when answering something that is not trained for in LNN.\\
% Forward/backward algorithms is based on proof rules. One can pull out these proof rules and see exactly why something is ... \\
% Integrate prior knowledge: you can include logical statements where you set the weights based on the hard value. Tailored activation functions give you the classical outputs and classical inputs so at least if everything you're dealing with where you have a constraints you know you're going to have something classical such as the airplane is in the air, yes or no?\\  
% To ensure reasonable settings of parameters, the authors show how the parameter-learning problem can be framed as an optimization problem with constraints (learning):
% \\
% several drawbacks
% \\
% So here come the tailored activation function to prevent constraints in the learning\\

% \subsection{NeurASP}
% \label{subsec:neurasp}
% Neuro answer set program. The key contribution is defining the backpropagation term\\

% No scalability because backpropagation has to compute against all the answer sets.\\
% weak supervision: we don't have the exact labels but we have information about the labels. 

\section{Representative Neuro-Symbolic Models}
\label{sec:model}

This section presents selected widely-used neuro-symbolic AI workloads as representative ones for our analysis. We consider them representative because they are diverse in terms of applications, model structures, and computational patterns. 

\begin{figure*}
% \vspace{-6pt}
\begin{minipage}[b]{\linewidth}
    \centering
    \includegraphics[width=\columnwidth]{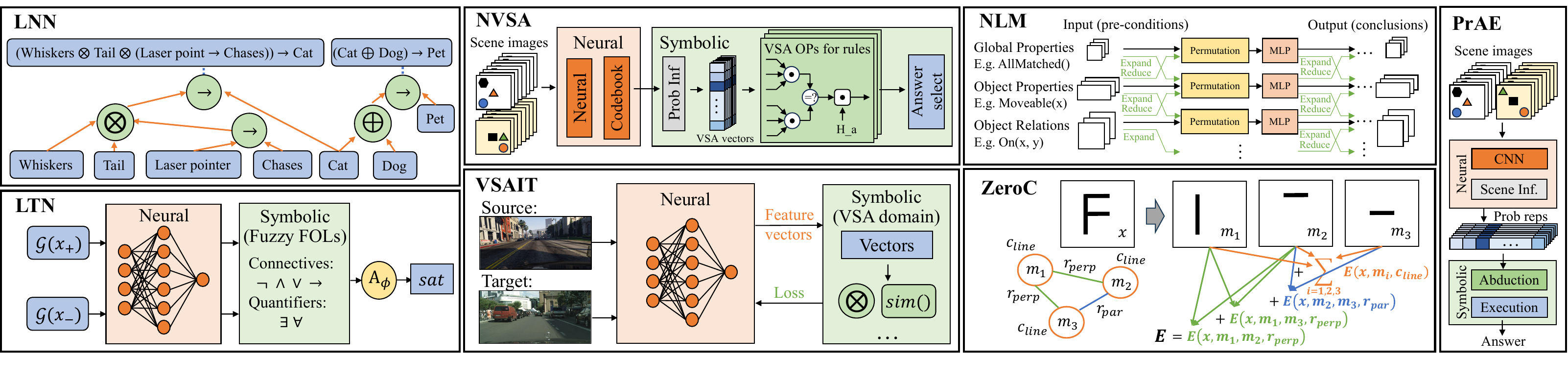}
\end{minipage}% 
 \vspace{-5pt}
\begin{minipage}[b]{\linewidth}
\Huge
    \renewcommand*{\arraystretch}{1.15}
    \setlength\tabcolsep{2pt}
\resizebox{\linewidth}{!}{%

% \begin{table*}[t!]

% \vspace{-5pt}
\centering
% \caption{xxx}
% \renewcommand*{\arraystretch}{1.05}
% \setlength\tabcolsep{2pt}
% \resizebox{\linewidth}{!}{%
\begin{tabular}{cc|c|c|c|c|c|c|c}
\hline
\multicolumn{2}{c|}{\textbf{\begin{tabular}[c]{@{}c@{}}Representative Neuro- \\ Symbolic AI Workloads\end{tabular}}}                                                                             & \textbf{\begin{tabular}[c]{@{}c@{}}Logic Neural \\ Network~\cite{riegel2020logical}\end{tabular}}                                                                                                          & \textbf{\begin{tabular}[c]{@{}c@{}}Logic Tensor \\ Network~\cite{badreddine2022logic}\end{tabular}}                                                                                                                                                                                   & \textbf{\begin{tabular}[c]{@{}c@{}}Neuro-Vector-Symbolic \\ Architecture~\cite{hersche2023neuro}\end{tabular}}                                                                                               & \textbf{\begin{tabular}[c]{@{}c@{}}Neural Logic \\ Machine~\cite{dongneural}\end{tabular}}                                       & \textbf{\begin{tabular}[c]{@{}c@{}}Vector Symbolic Architecture \\ Image2Image Translation~\cite{theiss2022unpaired}\end{tabular}}                                                         & \textbf{\begin{tabular}[c]{@{}c@{}}Zero-shot Concept Recog-\\ nition and Acquisition~\cite{wu2022zeroc}\end{tabular}}                                                                                           & \textbf{\begin{tabular}[c]{@{}c@{}}Probabilistic Abduction\\ and Execution~\cite{zhang2021abstract}\end{tabular}}                                                                                                                                                          \\ \hline
\multicolumn{2}{c|}{\textbf{Abbreviation}}                                                                                                                                                        & LNN                                                                                                                                                                               & LTN                                                                                                                                                                                                                                                        & NVSA                                                                                                                                                                                 & NLM                                                                                                            & VSAIT                                                                                                                                                                     & ZeroC                                                                                                                                                                                       & PrAE                                                                                                                                                                                                                                               \\ \hline
\multicolumn{2}{c|}{\textbf{Neuro-Symbolic Category}}                                                                                                                                           & Neuro:Symbolic$\rightarrow$Neuro                                                                                                                                                              & $\mbox{Neuro}_{\mbox{Symbolic}}$                                                                                                                                                                                                                                            & Neuro$|$Symbolic                                                                                                                                                                     & Neuro{[}Symbolic{]}                                                                                            & Neuro$|$Symbolic                                                                                                                                                          & Neuro{[}Symbolic{]}                                                                                                                                                                         & Neuro$|$Symbolic                                                                                                                                                                                                                                   \\ \hline
\multicolumn{2}{c|}{\textbf{Learning Approach}}                                                                                                                                                 & Supervised                                                                                                                                                                        & \begin{tabular}[c]{@{}c@{}}Supervised/Unsupervised \end{tabular}                                                                                                                                                                                        & \begin{tabular}[c]{@{}c@{}}Supervised/Unsupervised\end{tabular}                                                                                                                  & \begin{tabular}[c]{@{}c@{}}Supervised/Unsupervised\end{tabular}                                            & Supervised                                                                                                                                                 & Supervised                                                                                                                                                                                  & \begin{tabular}[c]{@{}c@{}}Supervised/Unsupervised\end{tabular}                                                                                                                                                                                \\ \hline
\multicolumn{1}{c|}{\multirow{6}{*}{\textbf{\begin{tabular}[c]{@{}c@{}}Deployment\\ Scenario\end{tabular}}}}  & \textbf{Application}                                                            & \begin{tabular}[c]{@{}c@{}}Learning and reasoning,\\ Full theorem prover\end{tabular}                                                                                             & \begin{tabular}[c]{@{}c@{}}Querying, learning, reasoning\\ (relational and embedding\\ learning, query answering)\end{tabular}                                & \begin{tabular}[c]{@{}c@{}}Fluid intelligence,\\ Abstract reasoning\end{tabular}                                                                                                     & \begin{tabular}[c]{@{}c@{}}Relational reasoning,\\ Decision making\end{tabular}                                & \begin{tabular}[c]{@{}c@{}}Unpaired image-to-image \\ translation\end{tabular}   & \begin{tabular}[c]{@{}c@{}}Cross-domain classification\\ and detection, Concept \\ acquisition\end{tabular}                                                           & \begin{tabular}[c]{@{}c@{}}Fluid intelligence,\\ Spatial-temporal reasoning\end{tabular}                                                                                                                                                           \\ \cline{2-9} 
\multicolumn{1}{c|}{}                                                                                         & \textbf{\begin{tabular}[c]{@{}c@{}}Advantage vs. \\ Neural Model\end{tabular}} & \begin{tabular}[c]{@{}c@{}}Higher interoperability, \\ resilience to incomplete  \\knowledge,  generalization\end{tabular} & \begin{tabular}[c]{@{}c@{}}Higher data efficiency,\\ comprehensibility, out-of-\\distribution generalization \end{tabular}                                                                                                                                          & \begin{tabular}[c]{@{}c@{}} Higher joint representations\\ efficiency, abstract reasoning \\ capability, transparency\end{tabular} & \begin{tabular}[c]{@{}c@{}} Higher generalization, \\ logic reasoning, deduction, \\ explainability capability\end{tabular}                                                                               & \begin{tabular}[c]{@{}c@{}}Address semantic flipping and \\ hallucinations issue in unpaired \\ image translation tasks\end{tabular}                        & \begin{tabular}[c]{@{}c@{}}Higher generalization, concept\\ acquisition and recognition,\\ compositionality capability\end{tabular} & \begin{tabular}[c]{@{}c@{}}Higher generalization,\\ transparency, interpre-\\ tability, and robustness\end{tabular} \\ \cline{2-9} 
\multicolumn{1}{c|}{}                                                                                         & \textbf{Dataset}                                                                & \begin{tabular}[c]{@{}c@{}}LUBM benchmark~\cite{guo2005lubm},\\ TPTP benchmark~\cite{sutcliffe2017tptp}\end{tabular}                                              & \begin{tabular}[c]{@{}c@{}}UCI~\cite{asuncion2007uci}, Leptograpsus \\ crabs~\cite{gencer2023research}, DeepProbLog~\cite{manhaeve2018deepproblog}\end{tabular} & \begin{tabular}[c]{@{}c@{}}RAVEN~\cite{zhang2019raven}, \\ I-RAVEN~\cite{hu2021stratified}, PGM~\cite{pgm} \end{tabular}                                                                                                                                                   & \begin{tabular}[c]{@{}c@{}}Family graph reasoning,\\ sorting, path finding~\cite{graves2016hybrid}\end{tabular} & \begin{tabular}[c]{@{}c@{}}GTA~\cite{richter2016playing}, Cityscapes~\cite{cordts2016cityscapes}, \\ Google Maps dataset~\cite{isola2017image}\end{tabular} & \begin{tabular}[c]{@{}c@{}}Abstraction reasoning~\cite{chollet2019measure},\\ Hierarchical-concept corpus~\cite{shanahan2020explicitly}\end{tabular}                                                       & \begin{tabular}[c]{@{}c@{}}RAVEN~\cite{zhang2019raven}, \\ I-RAVEN~\cite{hu2021stratified}, PGM~\cite{pgm} \end{tabular}                                                                                                                                                                                                                             \\ \hline
\multicolumn{1}{c|}{\multirow{3}{*}{\textbf{\begin{tabular}[c]{@{}c@{}}Computation \\ Pattern\end{tabular}}}} & \textbf{Datatype}                                                               & FP32                                                                                                                                                                              & FP32                                                                                                                                                                                                                                                       & FP32                                                                                                                                                                                 & FP32                                                                                                           & FP32                                                                                                                                                                      & INT64                                                                                                                                                                                       & FP32                                                                                                                                                                                                                                               \\ \cline{2-9} 
\multicolumn{1}{c|}{}                                                                                         & \textbf{Neural}                                                              & Graph                                                                                                                                                                             & MLP                                                                                                                                                                                                                                                        & ConvNet                                                                                                                                                                              & Sequential tensor                                                                                               & ConvNet                                                                                                                                                                   & Energy-based network                                                                                                                                                                        & ConvNet                                                                                                                                                                                                                                            \\ \cline{2-9} 
\multicolumn{1}{c|}{}                                                                                         & \textbf{Symbolic}                                                           & FOL/Logical operation                                                                                                                                                  & FOL/Logical operation                                                                                                                                                                                                                           & VSA/Vector operation                                                                                                                                                          & FOL/Logical operation                                                                                              & VSA/Vector operation                                                                                                                                                           & Graph, vector operation                                                                                                                                                                           & VSA/Vector operation                                                                                                                                                                                                                                             \\ \hline
\end{tabular}
}
\captionof{table}{Selected neuro-symbolic AI workloads for analysis, representing a diverse of categories, applications, and computational patterns.}
\vspace{-10pt}
\label{tab:selected_model}
\end{minipage}
\end{figure*}

\vspace{-0.1in}
\subsection{Model Overview.}
We select seven neuro-symbolic AI models for profiling analysis (Tab.~\ref{tab:selected_model}): LNN on logic program tasks~\cite{riegel2020logical}, LTN on querying and reasoning tasks~\cite{badreddine2022logic}, NVSA~\cite{hersche2023neuro} on the Raven's Progressive Matrices task~\cite{zhang2019raven}, NLM on relational reasoning and decision making tasks~\cite{dongneural}, VSAIT on unpaired image-to-image translation tasks~\cite{theiss2022unpaired}, ZeroC on cross-domain classification and detection tasks~\cite{wu2022zeroc}, and PrAE on spatial-temporal reasoning tasks~\cite{zhang2021abstract}. These selected workloads represent Neuro:Symbolic$\rightarrow$Neuro, $\mbox{Neuro}_{\mbox{Symbolic}}$, Neuro$|$Symbolic, and Neuro[Symbolic] systems (Sec.~\ref{sec:survey}), respectively. Interested readers could refer to their references for more details.

% \vspace{/-0.1in}
\subsection{Logical Neural Network (LNN)}
LNN is a neuro-symbolic framework that integrates neural learning with symbolic logic, enabling direct interpretability, domain knowledge utilization, and robust problem-solving~\cite{riegel2020logical}. LNNs map neurons to logical formula elements, using parameterized functions to represent logical connectives (e.g., $\wedge, \vee$) with constraints to preserve logical behavior. By combining facts and rules within a neural framework, LNNs use weighted real-valued logics via Łukasiewicz logic~\cite{badreddine2022logic}. Compared to neural models, LNNs offer superior logical expressivity, tolerance to incomplete knowledge, and general task applicability, excelling in theorem proving with compositional, modular structures.

\vspace{-0.1in}
\subsection{Logical Tensor Network (LTN)}
LTN is a neuro-symbolic framework for querying, learning, and reasoning with data and abstract knowledge using fuzzy first-order logic (FOL)~\cite{badreddine2022logic}. LTN grounds FOL elements in data using neural graphs and fuzzy logic, transforming connectives into real values and interpreting quantifiers via approximate aggregations~\cite{badreddine2022logic}. The network computes truth degrees using embedded tensor representations. Compared to neural models, LTN enhances explainability, data efficiency, and out-of-distribution generalization by expressing knowledge through logical axioms over data.

\vspace{-0.1in}
\subsection{Neuro-Vector-Symbolic Architecture (NVSA)}
NVSA is a neuro-symbolic architecture for abstract reasoning, combining neural visual perception and vector-symbolic probabilistic reasoning to improve abduction reasoning efficiency~\cite{hersche2023neuro}. NVSA uses holographic distributed representations to co-design visual perception and probabilistic reasoning, enabling perceptual representations and symbolic rule processing for accurate Raven's progressive matrices (RPM)~\cite{raven,iraven} test performance.
Compared to neural models, NVSA overcomes the binding problem and superposition catastrophe, achieving superior accuracy in RPM tests and even surpassing human performance.

\vspace{-0.1in}
\subsection{Neural Logic Machine (NLM)}
NLM is a neuro-symbolic architecture for inductive learning and logical reasoning, combining neural networks as function approximators with logic programming for symbolic processing~\cite{dongneural}.
NLM approximates logic operations using neural networks and implements logic quantifiers through neural module wiring. Its multi-layer structure deduces object relations, forming higher abstractions with increased layers.
Compared to neural models, NLM excels in relational reasoning and decision-making, generalizing well from small-scale to large-scale tasks, outperforming traditional neural networks and logic programming.

\vspace{-0.1in}
\subsection{Vector Symbolic Architecture-Based Image-to-Image Translation (VSAIT)}
VSAIT addresses semantic flipping in image translation between domains with large distribution gaps, leveraging vector-symbolic architecture for photorealism and robustness~\cite{theiss2022unpaired}.
VSAIT learns invertible mappings in hypervector space, ensuring consistency between source and translated images while encoding features into random vector-symbolic hyperspace.
Compared to neural models, VSAIT ensures robustness to semantic flipping and significantly reduces image hallucinations observed for unpaired image translation between domains with large gaps.

\vspace{-0.1in}
\subsection{Zero-Shot Concept Recognition and Acquisition (ZeroC)}
ZeroC is a neuro-symbolic architecture that recognizes and acquires novel concepts in a zero-shot manner by leveraging symbolic graph structures~\cite{wu2022zeroc}.
ZeroC uses graphs and energy-based models to represent concepts and relations, allowing hierarchical concept models to generalize across domains during inference.
Compared to neural models, ZeroC excels in zero-shot concept recognition, surpassing neural models in tasks requiring novel concept learning without extensive examples.

\vspace{-0.1in}
\subsection{Probabilistic Abduction and Execution (PrAE) Learner}
PrAE is a neuro-symbolic learner for spatial-temporal cognitive reasoning, centered on probabilistic abduction and execution of scene representations~\cite{zhang2021abstract}.
PrAE combines neural visual perception with symbolic reasoning to predict object attributes and generate probabilistic scene representations, inferring hidden rules for systematic generalization.
Compared to neural models, PrAE outperforms them in spatial-temporal reasoning, offering transparency, interpretability, and human-level generalizability.

\section{Workload Characterization Methodology}
% \vspace{-0.3em}
\label{sec:method}
% In this section, we analyze the performance characteristics of three recent NSAI models and discuss the system bottleneck, workload operators, and optimization opportunities.
This section presents our neuro-symbolic AI workload profiling methodology (Sec.~\ref{subsec:profile_method}) and operator characterization taxonomy (Sec.~\ref{subsec:taxonomy}) that will be leveraged in Sec.~\ref{sec:results}.

\vspace{-0.1in}
\subsection{Workload Profiling Methodology} 
\label{subsec:profile_method}
We first conduct function-level profiling to capture statistics such as runtime, memory, invocation counts, tensor sizes, and sparsity of each model, by leveraging the built-in PyTorch Profiler. We also perform post-processing to partition the characterization results into various operation categories. The experiments are conducted on a system with Intel Xeon Silver 4114 CPU and Nvidia RTX 2080 Ti GPU (250W), as well as edge SoCs such as Xavier NX (20W) and Jetson TX2 (15W).

% , and then use the PyTorch Profiler~\cite{pytorch_profiler} to measure the CPU and GPU runtimes of each model at a per-function granularity. The experiments are conducted on a system with an Intel Xeon Silver 4114 CPU and an Nvidia RTX 2080 Ti GPU.

% Need: summarize evaluation metrics (runtime, memory, operator, sparsity, etc)

% Need: a table to compare different computing platforms.
% \textbf{Compute Operator Analysis Method.} We then perform compute operator-level profiling for further analysis. We classify each neuro and symbolic workload of the LNN, LTN, and NVSA model into six operator categories: convolution, matrix multiplication (MatMul), vector/element-wise operation (e.g., tensor add, div, norm), data transformation (e.g., reshape, transpose), data movement (e.g., inter-device transfer), and others (e.g., fuzzy logic, logic rule). 

\begin{figure*}[t!]
\vspace{-0.02in}
\centering\includegraphics[width=2.05\columnwidth]{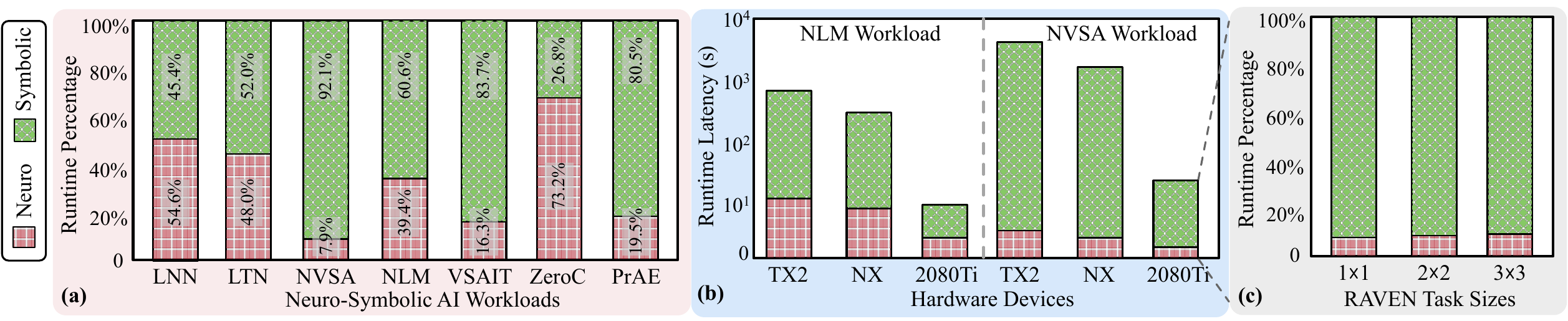}
\vspace{-0.1in}
        \caption{\textbf{Neural and symbolic runtime latency characterization.} \textbf{(a)} Benchmark seven representative neuro-symbolic workloads (LNN, LTN, NVSA, NLM, VSAIT, ZeroC, PrAE) on the CPU+GPU system, showing symbolic may serve as system bottleneck. \textbf{(b)} Benchmark NVSA and NLM workloads on Jetson TX2, Xavier NX, and RTX GPU, showing that real-time performance cannot be satisfied. \textbf{(c)} Benchmark NVSA workload on various RPM task sizes on RTX GPU, indicating the potential scalability problem and consistent symbolic bottleneck.}
        \label{fig:profiling_latency}
        \vspace{-10pt}
\end{figure*}

\vspace{-0.1in}
\subsection{Workload Characterization Taxonomy} 
\label{subsec:taxonomy}
On top of function-level profiling, we further conduct compute operator-level profiling for further analysis. We classify each neural and symbolic workload of the LNN, LTN, NVSA, NLM, VSAIT, ZeroC, and PrAE neuro-symbolic models into six operator categories: convolution, matrix multiplication (MatMul), vector/element-wise tensor operation, data transformation, data movement, and others~\cite{susskind2021neuro}. 

\textbf{Convolution}: refers to operations involving overlaying a matrix (kernel) onto another matrix (input) and computing the sum of element-wise products. This process is slid across the entire matrix and transforms the data. Convolution is common in neural networks and leads to high operational intensity. 

\textbf{Matrix Multiplication}: refers to general matrix multiplication (GEMM) with two matrices, either dense or sparse. Fully-connected layers in neural networks use GEMM as their primary mathematical operation. Multiplication of large, dense matrices is typically computationally intensive but highly parallelizable. There is typically a trade-off between the generality of the sparsity and the overhead of hardware optimization. Sparse matrix multiplication requires efficient mechanisms to perform lookups into the tables of non-zero values.

% \textbf{Dense Matrix Multiplication}: refers to general matrix multiplication (GEMM) with two dense matrices. Fully-connected layers in neural networks use GEMM as their primary mathematical operation. Multiplication of large, dense matrices is typically computationally intensive but highly parallelizable. 

% \textbf{Sparse Matrix Multiplication}: refers to GEMM operations with sparse matrices where most of the elements are zero. There is typically a trade-off between the generality of the sparsity and the overhead of hardware optimization. Sparse matrix multiplication requires efficient mechanisms to perform lookups into the tables of non-zero values.

\textbf{Vector/Element-wise Tensor Operation}: refers to operations performed element-wise on tensors (generalized matrices, vectors, and higher-dimensional arrays), including addition, subtraction, multiplication, and division, applied between two tensors element by element, as well as activation, normalization, and relational operations in neuron models.

\textbf{Data Transformation}: refers to operations that reshape or subsample data, including matrix transposes, tensor reordering, masked selection, and coalescing which is a process in which duplicate entries for the same coordinates in a sparse matrix are eliminated by summing their associated values.

\textbf{Data Movement}: refers to data transferring from memory-to-compute, host-to-device, and device-to-host, as well as operations such as tensor duplication and assignment.
% Many types of operations require data movement but little or no computation, primarily consisting of

\textbf{Others}: refers to operations such as fuzzy first of logic and logical rules that are utilized in some symbolic AI workloads.

\begin{figure*}[t!]
\vspace{-0.05in}
\centering\includegraphics[width=2.05\columnwidth]{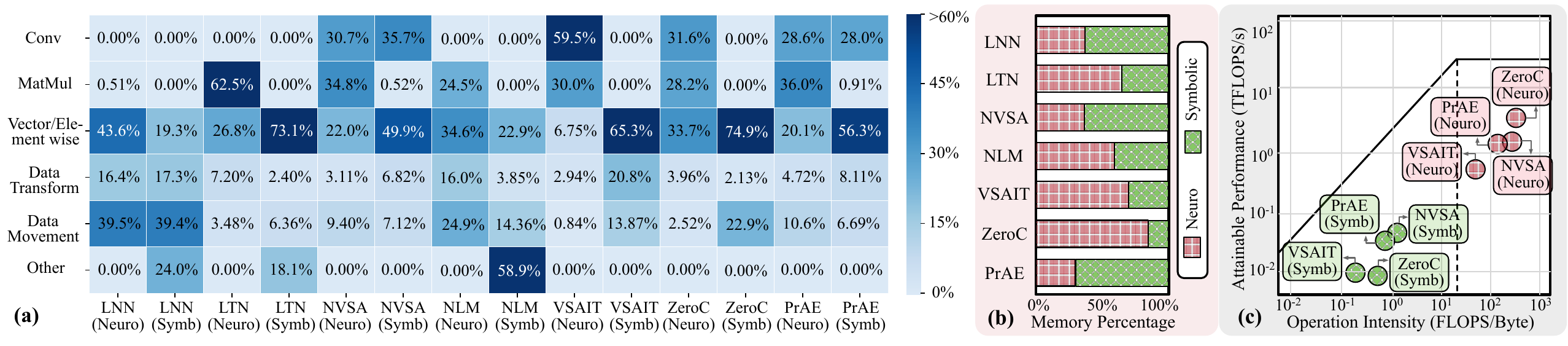}
\vspace{-0.15in}
        \caption{\textbf{Compute operators, memory and roofline characterization.} \textbf{(a)} Compute operator runtime ratio of representative neuro-symbolic workloads, indicating neural operations mainly consisting of MatMul and Conv, while symbolic operations with vector/tensors. \textbf{(b)} Benchmark memory usage during computation and \textbf{(c)} roofline analysis on RTX 2080Ti GPU, showing typically neural operations are compute-bounded and symbolic operations are memory-bounded.}
        \label{fig:profiling_roofline}
        \vspace{-10pt}
\end{figure*}

% \vspace{-0.5em}
\section{Workload Characterization Results}
\label{sec:results}
% \blue{add GPU results; 
% Reviewer A: Is high-end GPU a target for these neurosymbolic applications?; 
% Reviewer B: Will there be overlaps between neuro and symbolic phases?;
% }

% \blue{mention VSA in the takeaways, we take this further to provide hardware acceleration, connect with sec.VI}

% \vspace{-0.3em}
% \textbf{Runtime Breakdown.}
% Compared to neuro workloads, symbolic workloads are not negligible in compute latency and likely to pose a system bottleneck. Fig.~\ref{fig:operator}\textcolor{Blue}{(a)} shows the runtime breakdown for each model, where neuro (symbolic) workloads account for 54.6\% (45.4\%), 48.0\% (52.0\%), 7.9\% (92.1\%) runtime of LNN, LTN, and NVSA, respectively. Notably, symbolic workload dominates NVSA runtime, predominately due to the sequential and computational-intensive rule detection during the reasoning procedure. This reasoning computation depends on the result of frontend neuro workload and thus lies on the critical path for inference, however, there are still opportunities for data pre-processing and parallel rule query to reduce the bottleneck.
This section analyzes the performance characteristics of representative neuro-symbolic workloads and discusses their runtime and scalability (Sec.~\ref{subsec:latency}), compute operators (Sec.~\ref{subsec:operator}), memory usage (Sec.~\ref{subsec:memory_roofline}), operation graph (Sec.~\ref{subsec:graph}), hardware utilization (Sec.~\ref{subsec:hw_inefficiency}), and sparsity (Sec.~\ref{subsec:sparsity}).
% and optimization opportunities.

\vspace{-0.1in}
\subsection{Compute Latency Analysis}
\label{subsec:latency}
\textbf{End-to-end latency breakdown.}
We first characterize the end-to-end latency of representative neuro-symbolic AI workloads (Fig.~\ref{fig:profiling_latency}). We can observe that 
(1) Compared to neural workloads, symbolic workloads are not negligible in computing latency and may become a system bottleneck. For example, the neural (symbolic) workloads account for 54.6\% (45.4\%), 48.0\% (52.0\%), 7.9\% (92.1\%), 39.4\% (60.6\%), 16.3\% (83.7\%), 73.2\% (26.8\%), and 19.5\% (80.5\%) runtime of LNN, LTN, NVSA, NLM, VSAIT, ZeroC, and PrAE models, respectively (Fig.~\ref{fig:profiling_latency}\blue{a}). Notably, the symbolic workload dominates the NVSA's runtime, predominately due to the sequential and computational-intensive rule detection during the involved reasoning procedure. 
(2) The real-time performance cannot be satisfied, e.g., RTX 2080Ti GPU takes 380~s and TX2 takes 7507~s for RPM task in NVSA (Fig.~\ref{fig:profiling_latency}\blue{b}). Even if more computing resources are available to reduce neural inference time, the significant overhead of vector-symbolic-based reasoning still prohibits real-time execution.
(3) The symbolic operations may not be well accelerated by GPU. For example, symbolic counts for 92.1\% of total NVSA inference time while its floating-point operations (FLOPS) count for only 19\% of total FLOPS, indicating inefficient computation.
% (3) Symbolic backend reasoning computation depends on the result of the frontend neuro workload and thus lies on the critical path during inference. 

\colorbox{lightgray!50}{\textbf{Takeaway 1:}} \textit{Neuro-symbolic AI models typically exhibit high latency compared to neural models, prohibiting them from real-time applications. Symbolic operations are processed inefficiently on CPU/GPUs and may result in system bottlenecks.}
% Runtime latency: The total runtime is high, symbolic operation is not negligible, in some workload symbolic comparable with neuro workload and even result in system bottleneck. 
% Fig: show runtime scalability with problem complexity evolving.

\textbf{End-to-end latency scalability.} 
We evaluate the end-to-end runtime across various task sizes and complexities, as shown in Fig.~\ref{fig:profiling_latency}\textcolor{blue}{c} of RPM task for NVSA. We can observe that 
(1) The neural vs. symbolic runtime proportion remains relatively stable across various task sizes. For example, when task size increases from 2$\times$2 to 3$\times$3, the symbolic runtime slightly changes from 91.59\% to 87.35\%.
(2) The total runtime increases quadratically with task size evolving. For example, the total runtime increases 5.02$\times$ in the above case, indicating the potential scalability bottleneck of neuro-symbolic models.

% \textbf{Runtime Scalability.} We observe that the neuro vs. symbolic runtime proportion shown in Fig.~\ref{fig:operator}\textcolor{blue}{(a)} remains relatively stable across various test sets under the same size, whereas the total runtime increases quadratically with the test set size. For example, when the test set size increases from 2$\times$2 to 3$\times$3, the symbolic workload runtime percentage only increases from 92.06\% to 94.71\%, but the total runtime of the NVSA model increases by 5.02$\times$, indicating the potential scalability bottleneck of NSAI models.

% For NLM, it can be generalized to solve large-scale tasks by training on small-scale tasks. The rule complexity also increase, show scalability.

\colorbox{lightgray!50}{\textbf{Takeaway 2:}} \textit{The neural and symbolic components runtime ratio remains relatively stable while total latency explodes with the task complexity evolving. The potential scalability bottleneck calls for highly scalable and efficient architecture.}

\colorbox{brown!30}{\textbf{Recommendation 1:}} \textit{Optimization on neuro-symbolic workloads from algorithm-system-hardware cross-layer perspectives is highly desirable for achieving real-time, efficient and scalable cognitive systems.}

% \subsection{Workload Analysis}
% \textbf{Workload Operator Overview.} 
\subsection{Compute Operator Analysis}
\label{subsec:operator}
% Fig.~\ref{fig:operator}\textcolor{Blue}{(b)} partitions the neuro and symbolic workloads of LNN, LTN, and NVSA models into six operator categories with runtime breakdown. Below is the workload analysis:
Fig.~\ref{fig:profiling_roofline}\textcolor{blue}{a} partitions the neural and symbolic workloads of the LNN, LTN, NVSA, NLM, VSAIT, ZeroC, and PrAE workloads into six operator categories (Sec.~\ref{subsec:taxonomy}) with runtime latency breakdown. We make the following observations:

% \textbf{Neuro Workload Analysis.}
% The neuro workload is dominated by MatMul and activations. LTN (neuro) is dominated by MatMul due to its heavy MLP components. NVSA (neuro) spends the majority of its runtime on MatMul and convolution because it uses ResNet18 as the perception backbone for feature extraction. By contrast, LNN (neuro) spends a large portion of its runtime on element-wise operations due to the sparse syntax tree structure composed of proposition logic. Notably, data movement also takes up a significant amount of LNN (neuro) runtime because of its unique bidirectional dataflow during reasoning inference.
\textbf{Neural Workload Analysis.}
The neural workload is dominated by the MatMul and activation operations. LTN (neuro) is dominated by MatMul due to its heavy MLP components, while NVSA, VSAIT, and PrAE's (neuro) majority runtime is on MatMul and convolution because they adopt the neural network as the perception backbone for feature extraction. By contrast, a large portion of LNN and NLM's (neuro) runtime is on vector and element-wise tensor operations due to the sparse syntax tree structure composed of proposition logic and the sequential logic deduction computations on multi-group architecture. Notably, data movement also takes up a significant amount of LNN (neuro) runtime because of its unique bidirectional dataflow during reasoning inference.

% \textbf{Symbolic Workload Analysis.}
% The symbolic workload is dominated by vector and scalar operations that exhibit low operational intensities and complex control flow. Both LNN (symbolic) and LTN (symbolic) have a large number of logic operations that pose parallelism optimization opportunities in their database queries and arithmetic operations, especially for larger symbolic models. Moreover, LNN (symbolic) is severally data movement-bounded due to its sparse and irregular memory access and bidirectional inference, whereas model-aware dataflow architecture would likely be beneficial for relieving its data movement bottleneck.
% Notably, the element-wise operations usually stem from high-dimensional distributed vector computation (e.g., binding, bundling, permutation) for symbolic representation, which is difficult to process efficiently on GPUs.
% Therefore, the challenges of accelerating these element-wise low-operational-intensity computations will become increasingly important as the task and feature complexities further grow, and potentially leads us down a path toward NSAI systems that are as computationally efficient as brains.
\textbf{Symbolic Workload Analysis.}
The symbolic workload is dominated by vector and scalar operations that exhibit low operational intensities and complex control flows. Both LNN, LTN, and NLM's (symbolic) have a large number of logic operations, posing parallelism optimization opportunities in their database queries and arithmetic operations, especially for larger symbolic models. Meanwhile, LNN (symbolic) is severally data movement-bounded due to its sparse and irregular memory accesses and bidirectional inference, where model-aware dataflow architecture would likely be beneficial for alleviating this bottleneck. NVSA, VSAIT, and PrAE's (symbolic) are composed of vectors for vector-symbolic operations.
% Celine: it seems "element-wise operations" appear abruptly here
Notably, these operations usually stem from high-dimensional distributed vector computations (e.g., binding, bundling) for symbolic representation, which are difficult to process efficiently on GPUs.
Therefore, the challenges of accelerating these computations will become increasingly important as the task and feature complexities further grow. We leverage VSA kernels as a case study and present a cross-layer optimization solution in Sec.~\ref{sec:hardware} to improve system efficiency.
% , and potentially leads us down a path toward NSAI systems that are as computationally efficient as brains.

% \textbf{Runtime Scalability.} We observe that the runtime and distribution in Fig.~\ref{fig:operator}\textcolor{Blue}{(a)} roughly keep stable for various test cases with the same size. However, when the RAVEN test set size increases from 2$\times$2 to 3$\times$3, the total runtime of the NVSA model increases by 5.02$\times$ and symbolic workload runtime percentage increases from 92.06\% to 94.71\%, indicating the potential scalability bottleneck of NSAI models.

% Fig: confusion matrix to show the operator ratio of each workload

\colorbox{lightgray!50}{\textbf{Takeaway 3:}} \textit{The neural components mainly consist of MatMul and Convs, while the symbolic components are dominated by vector/element-wise tensor and logical operations. The data transfer overhead arising from the separate neural and symbolic execution on GPUs and CPUs poses efficient hardware design challenges.}
% Operator analysis: Symbolic involves lots of vector and element-wise operations, complex control that are inefficient in CPU, GPU

\colorbox{brown!30}{\textbf{Recommendation 2:}} \textit{From the architecture level, custom processing units can be built for efficient symbolic operations (e.g., high-dimensional distributed vectors, logical operation, graph, etc). For non-overlap neural and symbolic components, reconfigurable processing units supporting both neural and symbolic operations are recommended.}

\subsection{Memory and System Analysis}
\label{subsec:memory_roofline}
% Fig: memory ratio of neural and symbolic of each workload

\textbf{Memory Usage Analysis.}
Fig.~\ref{fig:profiling_roofline}\textcolor{blue}{b} characterizes the memory usage of the LNN, LTN, NVSA, NLM, VSAIT, ZeroC, and PrAE workloads during computation. We can observe that
(1) PrAE (symbolic) consumes a high ratio of memory due to its large number of vector operations depending on intermediate results and exhaustive symbolic search. NVSA (symbolic) slightly alleviates the vector-symbolic operation memory by leveraging probabilistic abduction reasoning. ZeroC (neuro) contains energy-based models and process images in a large ensemble thus taking much memory.
(2) In terms of storage footprint, neural weights and symbolic codebooks typically consume more storage. For example, neural network and holographic vector-inspired codebook account for $>$90\% memory footprint in NVSA, because NVSA neural frontend enables the expression of more object combinations than vector space dimensions, requiring the codebook to be large enough to contain all object combinations and ensure quasi-orthogonality. 

% Communication bandwidth
\textbf{System Roofline Analysis.} Fig.~\ref{fig:profiling_roofline}\textcolor{blue}{c} employs the roofline model to quantify the memory boundedness of RTX 2080Ti GPU versions of the selected workloads. We observe that the symbolic components are in the memory-bound area while neural components are in the compute-bound area. For example, NVSA and PrAE symbolic operations require streaming vector elements to circular convolution computing units, increasing the memory bandwidth pressure. Optimizing the compute dataflow and leveraging the scalable and reconfigurable processing element can help provide this bandwidth.

% \colorbox{lightgray!50}{\textbf{Takeaway 4:}}  \textit{Memory Footprint: neuro weights typically account for most memory storage, codebook in symbolic also consumes lots of memory}

% Fig: roofline analysis for different workload: show neural is compute-bound and symbolic is memory-bound

\colorbox{lightgray!50}{\textbf{Takeaway 4:}}  \textit{Symbolic operations are memory-bounded due to large element streaming for vector-symbolic operations. Neural operations are compute-bounded due to computational-intensive MatMul/Convs. Neural weights and vector codebooks typically account for most storage while symbolic components require large intermediate caching during computation.}
% Communication bandwidth: vector-symbolic operations, such as circular conv-based binding and boundling operations require streaming many elements, memory-bound operation increasing the memory bandwidth pressure

\colorbox{brown!30}{\textbf{Recommendation 3:}} \textit{From the algorithm level, model compression (e.g., quantization and pruning) and efficient factorization of neural and symbolic components can be used to reduce memory and data movement overhead without sacrificing cognitive reasoning accuracy.}

\colorbox{brown!30}{\textbf{Recommendation 4:}} \textit{From the technology level, emerging memories and in/near-memory computing can alleviate the memory-bounded symbolic operations and improve scalability, performance, and efficiency of neuro-symbolic systems.}

\begin{figure}[t!]
\centering
    \includegraphics[width=1\columnwidth]{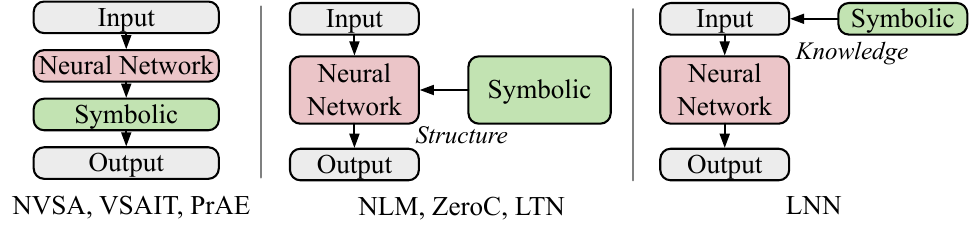}
        \centering
        \vspace{-0.15in}
        \caption{\textbf{Operator graph analysis.} Symbolic operation depends on neural results or needs to compile in neural structure as the critical path. Complex control and symbolic-only phase operation result in inefficiency and low hardware resource utilization.}
        \label{fig:graph}
        \vspace{-10pt}
\end{figure}

\subsection{Operation and Dataflow}
\label{subsec:graph}
% Fig: operator graph and hardware utilization
Fig.~\ref{fig:graph} analyzes the operation dependency in representative neuro-symbolic workloads. We can observe that the reasoning computation of NVSA, VSAIT, and PrAE depends on the result of the frontend neural workload and thus lies on the critical path during inference. LNN, LTN, NLM, and ZeroC need to compile the symbolic knowledge in neural representation or input embeddings. The complex control results in inefficiency in CPU and GPU, and the vector-symbolic computation period results in low hardware utilization. There are opportunities for data pre-processing, parallel rule query, and heterogeneous and reconfigurable hardware design to reduce this bottleneck.

\colorbox{lightgray!50}{\textbf{Takeaway 5:}}  \textit{The symbolic operations depend on the neural module results or need to compile into the neural structure, thus lying on the critical path of end-to-end neuro-symbolic systems. The vector-symbolic computation phase and complex control of neuro-symbolic components bring low hardware resource utilization and inefficiency in CPU/GPU.}

\colorbox{brown!30}{\textbf{Recommendation 5:}} \textit{From the system level, adaptive workload scheduling with parallelism processing of neural and symbolic components can be leveraged to alleviate resource underutilization and improve runtime efficiency.} 

% \colorbox{lightgray!50}{\textbf{Takeaway 8:}}  \textit{Utilization: vector-symbolic workloads exhibit low hardware resource utilization on off-the-shelf computing hardware}
\vspace{-0.1in}
\subsection{Hardware Inefficiency Analysis}
\label{subsec:hw_inefficiency}

The hardware inefficiencies of executing neuro-symbolic workloads mainly come from ALU underutilization, low cache hit rate, and massive data transfer. We leverage Nsight Systems/Compute tools to further characterize the GPU behavior of executing selected neuro-symbolic workloads. Tab.~\ref{tab:nsight_profile} lists the compute, memory, and data movement characteristics of representative neural and symbolic kernels in NVSA as an example.
We observe that typically in symbolic operations, the ALU utilization is $<$10\%, the L1 cache hit rate is around 20\%, the L2 cache hit rate is around 40\%, and DRAM bandwidth utilization is around 90\% with several memory-bounded. The data transfer memory operations account for around 50\% of total latency, where $>$80\% is from host CPU to GPU. Additionally, the synchronization overhead and waiting for GPU operations to complete results in CPU underutilization.

% Tab.~\ref{tab:nsight_profile} lists the compute, memory, and data movement characteristics of representative neural and symbolic kernels in NVSA as an example. We observe that symbolic kernels exhibit low ALU utilization, low cache hit rates, and high DRAM memory transactions, while neuro components exhibit high compute utilization. This further consolidates the inefficient computation of symbolic components on off-the-shelf CPU/GPU.

\begin{table}[t!]
\scriptsize
\centering
\caption{\textbf{Hardware inefficiency analysis.} \textnormal{The compute, memory, and communication characteristics of representative neural and symbolic kernels in NVSA workload executed on CPU/GPU platform.}}
% \vspace{0.1in}
\renewcommand*{\arraystretch}{1.05}
\setlength\tabcolsep{2.3pt}
\resizebox{\linewidth}{!}{%
\begin{tabular}{l|cc|cc}
\hline
\multirow{2}{*}{}        & \multicolumn{2}{c|}{\textbf{Neural Kernel}}                         & \multicolumn{2}{c}{\textbf{Symbolic Kernel}}                \\ \cline{2-5} 
                         & \multicolumn{1}{c|}{sgemm\_nn} & relu\_nn & \multicolumn{1}{c|}{vectorized\_elem} & elementwise \\ \hline
Compute Throughput (\%)  & \multicolumn{1}{c|}{95.1}              & 92.9             & \multicolumn{1}{c|}{3.0}              & 2.3         \\ \hline
ALU Utilization (\%)     & \multicolumn{1}{c|}{90.1}              & 48.3             & \multicolumn{1}{c|}{5.9}              & 4.5         \\ \hline
L1 Cache Throughput (\%) & \multicolumn{1}{c|}{79.7}              & 82.6             & \multicolumn{1}{c|}{28.4}             & 10.8        \\ 
L2 Cache Throughput (\%) & \multicolumn{1}{c|}{19.2}              & 17.5             & \multicolumn{1}{c|}{29.8}             & 22.8        \\ \hline
L1 Cache Hit Rate (\%)   & \multicolumn{1}{c|}{1.6}               & 51.6             & \multicolumn{1}{c|}{29.5}             & 33.3        \\
L2 Cache Hit Rate (\%)   & \multicolumn{1}{c|}{86.8}              & 65.5             & \multicolumn{1}{c|}{48.6}             & 34.3        \\ \hline
DRAM BW Utilization (\%) & \multicolumn{1}{c|}{14.9}              & 24.2             & \multicolumn{1}{c|}{90.9}             & 78.4        \\ \hline
\end{tabular}
}
\label{tab:nsight_profile}
% \vspace{-10pt}
\end{table}

\colorbox{lightgray!50}{\textbf{Takeaway 6:}}  \textit{While neural kernels exhibit high compute utilization and memory efficiency in GPUs, symbolic operations suffer from low ALU utilization, low L1 cache hit rates, and high memory transactions, resulting in low efficiency.}

\colorbox{brown!30}{\textbf{Recommendation 6:}} \textit{From the architecture level, heterogeneous or reconfigurable neural/symbolic architecture with efficient vector-symbolic units and high-bandwidth NoC can be optimized to improve ALU utilization and reduce data movement, thus improving system performance.}

\subsection{Sparsity Analysis}
\label{subsec:sparsity}
\begin{figure}[t!]
\centering
\includegraphics[width=\columnwidth]{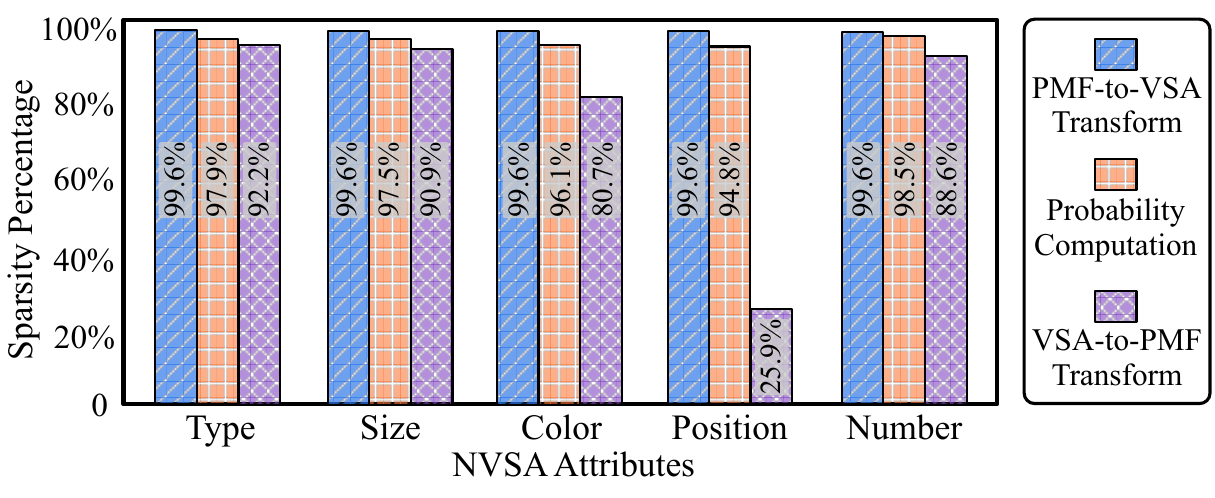}
        \centering
        \vspace{-0.15in}
        \caption{\textbf{Sparsity analysis.} The Sparsity ratio of NVSA symbolic operations, shows a high degree of sparsity with variations in attributes.}
        \label{fig:NVSA_sparsity}
        \vspace{-10pt}
\end{figure}

% Fig: show sparsity percentage of each workload
Neuro-symbolic workloads also exhibit sparsity features. For example, Fig.~\ref{fig:NVSA_sparsity} characterizes the sparsity of NVSA symbolic modules, including probabilistic mass function (PMF)-to-VSA transform, probability computation, and VSA-to-PMF transform, under different reasoning rule attributes. We can observe that NVSA has a high sparsity ratio ($>$95\%) with variations for specific attributes and unstructured patterns. Similarly, ZeroC and LNN also demonstrate $>$90\% sparsity ratio, while LTN features a dense computation pattern.

\colorbox{lightgray!50}{\textbf{Takeaway 7:}} \textit{Some neural and vector-symbolic components demonstrate a high level of unstructured sparsity with variations under different task scenarios and attributes.}

\colorbox{brown!30}{\textbf{Recommendation 7:}} \textit{From the algorithm and architecture level, sparsity-aware neural and symbolic algorithm and architecture design can benefit memory footprint, communication overhead, and computation FLOPS reduction.}
% Sparsity is observed in several NSAI models, including NVSA~\cite{hersche2023neuro}, LNN~\cite{riegel2020logical, sen2022neuro}, and CC (investigate others. I investigate LTN and NO).

% \begin{figure}[b!]
% \centering
% % \vspace{-1.7em}
%         \includegraphics[width=1.02\columnwidth]{Figs/breakdown_operator.pdf}
%         \centering
%         % \vspace{-2em}
%         \caption{(a) Runtime breakdown and (b) compute operation analysis for three neuro-symbolic AI models - LNN, LTN and NVSA. \textcolor{blue}{to remove}}
%         \label{fig:operator}
% \end{figure}

\vspace{-0.1in}
\subsection{Uniqueness of Neuro-Symbolic vs. Neural Networks}
\label{subsec:profile_summary}
To summarize, based on above analysis, neuro-symbolic AI workloads differ from neural networks mainly in three aspects:

\textbf{Compute kernels.} Neuro-symbolic workloads consist of heterogeneous neural and symbolic kernels. The symbolic operators (e.g., vector, graph, logic) are processed inefficiently on off-the-shelf CPUs/GPUs with low hardware utilization and cache hit and may result in runtime latency bottleneck.

\textbf{Memory.} Symbolic operations are memory-bounded due to large element streaming for vector-symbolic operations. Symbolic codebooks typically account for large memory footprints and require large intermediate caching during computation.

\textbf{Dataflow and scalability.} Neuro-symbolic workloads exhibit more complex control than NNs. Symbolic operations either critically depend on or compile in neural kernels. Their irregular dataflow, data dependency, and sequential processing bring low parallelism scalability and inefficiency in CPU/GPU.

% mention the large overhead of CPU-GPU communication in LTN.
% data movement 
% xx is dominated by xxx
% xx spends the majority of its runtime on xx; spends the most execution time on xx; a large portion
% a substantial portion of the remaining runtime is spent on xx

% \subsection{Application Performance Quality}
% Fig: show accuracy advantage of neurosymbolic workloads over NN models

% \colorbox{lightgray!50}{\textbf{Takeaway 1:}} \textit{Application performance: Neurosymbolic workload demonstrates improved cognition, transparency, data efficiency, and reasoning capability compared to conventional NN models, serving as a promising computing paradigm towards next-generation cognitive AI systems.}

% \subsection{Summary}

% For each workload, explain the reason behind the profiling data (shed light from neurosymbolic perspectives)
\section{Case Study: Hardware Acceleration of Vector-Symbolic Architecture}
\label{sec:hardware}
% \blue{CK: It gives me a feeling that we need to bridge this section more deeply with the previous profiling sections}
%\blue{Add a transition paragraph/subsection from profiling insight to hardware acceleration, and mention use VSA as a case study to show co-design; tie VSA with NVSA workflow; Two-level transition: profiling - hardware; VSA kernel - operator.}

\begin{table}[b!]
\scriptsize
\centering
\caption{\textbf{Design Features.} Features of the proposed VSA processor and their association with design recommendations from Sec.~\ref{sec:results}.}
\renewcommand*{\arraystretch}{1}
\setlength\tabcolsep{2.5pt}
\resizebox{1\columnwidth}{!}{
\begin{tabular}{l|c}
\hline
\textbf{VSA Processor Feature}&\textbf{Fulfilled Recommendation}\\ \hline
Compressed Storage of Symbols & Recommendation 3 (Sec.~\ref{subsec:memory_roofline}) \\ \hline
Distributed Memory System&Recommendation 4 (Sec.~\ref{subsec:memory_roofline}) \\ \hline
SIMD Multi-Tile Dataflow&Recommendation 5 (Sec.~\ref{subsec:graph}) \\ \hline
Heterogeneous Arithmetic Processing&Recommendation 6 (Sec.~\ref{subsec:hw_inefficiency}) \\ \hline
\end{tabular}}
\label{tab:hpu_features}
\end{table}

%Inspired by neuro-symbolic workload insights from the previous characterization study (Sec.~\ref{sec:results}), 
This section presents a cross-layer acceleration case study for vector-symbolic architecture (VSA), which 
%has been seamlessly merged and 
is a powerful model in many neuro-symbolic tasks~\cite{hersche2023neuro,theiss2022unpaired,kleyko2021vector,frady2020resonator}.
% is one of the essential components of neuro-symbolic systems. \blue{CK: VSA has been seamlessly merged and served as a sub-system in many neuro-symbolic tasks, including [help me reference].} 
%Building on the compute and memory system implications, 
We develop a design method consisting of accelerated vector-symbolic kernel formulation (Sec.~\ref{subsec:vsa_ops},~\ref{subsec:vsa_kernel}), architecture and dataflow (Sec.~\ref{subsec:accelerator_dataflow}), and programming method (Sec.~\ref{subsec:accelerator_control}), that overcomes computational inefficiencies from executing VSA components on CPUs and GPUs (Sec.~\ref{subsec:eval}).

Our proposed hardware design is inspired by neuro-symbolic workload insights from the characterization study in Sec.~\ref{sec:results}. Specifically, as shown in Tab.~\ref{tab:hpu_features}, it features (1) an energy-efficient dataflow with heterogeneous arithmetic units that can flexibly execute key vector-symbolic operations, (2) a distributed memory system employing near-memory computing to enhance scalability and memory performance, (3) compressed storage of symbolic operators to reduce the memory footprint of vector codebooks, and (4) a tiled design for vector-symbolic units to minimize data movement and optimize computational efficiency. These features collectively enable a highly efficient and scalable vector-symbolic hardware accelerator that significantly outperforms traditional platforms.
%\blue{MI: I think we may use a table here to match the above four characteristics with the profiling results/recommendations -- this could serve better in connecting Sec. VI with the previous ones.}

% We next present an overview of the accelerated vector-symbolic operations (Sec.~\ref{subsec:vsa_ops}), the accelerator's architecture (Sec.~\ref{subsec:accelerator_dataflow}), and its programming methodology (Sec.~\ref{subsec:accelerator_control}). 
\vspace{-0.1in}
\subsection{Vector-Symbolic Operations}
\label{subsec:vsa_ops}
%\blue{binding/bunding/distance/etc, profile at kernel-level, not operator-level, to design hardware for VSA, we need to understand VSA operator}

In the vector-symbolic kernel, computational elements, such as scalars and objects, are represented with hypervectors which can be manipulated by a set of algebraic operations~\cite{ibrahim2024efficient,wan2024h3dfact}, specifically, (1) binding, or element-wise multiplication, which creates a new hypervector that is quasi-orthogonal (dissimilar) to its constituents; (2) bundling, or element-wise addition, which combines hypervectors using element-wise majority count; (3) permutation, which rearranges the elements of a hypervector to preserve its order within a sequence; (4) scalar multiplication, which scales hypervector elements with a scalar weight 
% \blue{CK: I think VSA operations are not limited to the above four. I'm afraid the review might ask: how about XXX operation? which has been also found predominantly in XXX task. XXX task.}.
The similarity between vectors is measured using a variety of distance metrics, such as the dot product, Hamming distance, L1, and L2~\cite{xu2024ferex,shou2023see}. These operations collectively form a mathematical framework for implementing various cognitive functions tailored for VSA operations~\cite{kleyko2022vector}.
 % including the backend of NVSA

\vspace{-0.1in}
\subsection{Vector-Symbolic Kernel Formulation}
\label{subsec:vsa_kernel}

We present a description of operations and programmability features of our proposed hardware accelerator using a formal representation, i.e., kernel function. We express this kernel function as $O:=F(y,s)$, where $F(\cdot)$ integrates an array of kernel sub-functions $f_i$ that together cover the whole domain of accelerator operations, and $y=\{y_1, y_2,\ldots\}$ represents an array combining all item and prototype vectors used in computation. 
%The kernel sub-functions $F(\cdot)$ are distributed among $m$ computational elements $\{f_1,\ldots,f_m\}$, where each element implements a specific sequence of HD operations. 
The argument $s$ is defined by a group of conditional variables $s=(s_1, s_2, \ldots)$, which together are used to draw the sub-domains associated with the sub-functions $f_i$. 

The kernel functionality integrates computations for encoding and decoding, memory, and reasoning. Next, we formulate sub-functions $f_i$ to describe these computations. 

% \noindent
\textbf{Encoding and Decoding Kernel.} 
To facilitate the encoding and decoding, the kernel function needs to allow for flexible configuration of hypervector operations (binding, bundling, permutation). We take into account that binding can be distributed over bundling~\cite{kanerva2010we}, and propose the kernel function:

{\footnotesize
\[ a(y, (s_1, s_2)):=
    \begin{cases} 
      b\big(y, (s_2)\big); & s_1=0 \\
      \sum_i \big[b\big(y_i, (s_2)\big)\big]; & s_1=1
   \end{cases} ~~ \forall\{i,j\}\subset \mathbb{N} \] 
\[ b\big(y, (s_2)\big) :=
    \begin{cases} 
      y; & s_2=0 \\
      \Motimes_j (y_{j}); & s_2=1 \\
      \rho_j (y_j); & s_2=2 \\
      \Motimes_j\rho_{(j-1)}(y_{j}); & s_2=3
    \end{cases} ~~~~~~~~ \forall\{i,j\}\subset \mathbb{N} \]
% \[ \forall\{i,j\}\subset \mathbb{N} \]    
}

\noindent where $\rho_j$ means that the permutation operation ($\rho$) is repeated $j$ times, i.e., $\rho_3(x)=\rho(\rho(\rho(x)))$. Likewise, when $j=3$, the term $\Motimes_j(x_j)$ becomes equivalent to $(x_1\otimes x_2\otimes x_3)$, and also $\Motimes_j \rho_{(j-1)}(x_{j})$ becomes equivalent to $(x_1\otimes\rho(x_2)\otimes\rho(\rho(x_3)))$. 

% \noindent
\textbf{Resonator-Network Kernel.} 
This is a template VSA kernel for reasoning functions. Specifically, it takes as input a composed vector (which may represent a visual scene involving multiple objects as in the RPM problem) and seeks to factorize the vector into its constituent factors. The operation of the resonator network involves iterative steps for similarity evaluation and projection~\cite{frady2020resonator}. The kernel function used for projection can be defined as follows: $c(y):= \sum_i [n_i\times y_i]; \forall i\in \mathbb{N};\ n_i\in\mathbb{Z}$.
Here, $c(y)$ calculates a weighted sum of the vectors in $y$. The weight $n_i$ is given by a function $d(y_i, \bar{y})$, which measures the similarity between items $y_i$ and an estimate vector $\bar{y}\in y$.

% \noindent
\textbf{Nearest-Neighbor Search Kernel.} The similarity function $d(y_i, \bar{y})$ serves as the basis for identifying the closest vector to a query $\bar{y}\in y$ among an array of vectors $y=\{y_1,y_2,\cdots\}$. The array $y$ represents item vectors when performing a clean-up memory search, and prototype vectors when performing an associative memory search. To describe this operation, we use a kernel function defined as follows: $e(y) := \text{argmax}_i\ d(y_i, \bar{y})$.

% \noindent
\textbf{Kernel Support for Extended Vector Dimensions via Time-Multiplexing:} 
A particular advantage of element-wise vector operations (binding, bundling, and permutation) is that they can process full-scale vectors and time-multiplexed folds similarly. In contrast, similarity operations in $d(y_i, \bar{y})$ require the time-multiplexed folds to be collapsed into a single vector representation. When a similarity quantity is computed using only a single fold, it represents a partial quantity. Therefore, to obtain the total similarity value, $d(y_i, \bar{y})$ needs to aggregate these partial quantities. We express this condition as follows: $d(y_i, \bar{y}) := \sum_k \big(y_{ik}\cdot \bar{y}_k\big)\ \forall \ i\in\mathbb{N};\ k\in\{1,2,\ldots,L\}$.
Here, $L$ is the number of folds, and $\bar{y}_k$ and $y_{ik}$ are the $k$-th folds of the vectors $\bar{y}$ and $y_{i}$, respectively. The dot product measures the similarity between these folds, and the sum over all $k$ aggregates the similarities computed for all the folds.

% \noindent
\textbf{Compact Kernel Formalism.} Considering the information presented above, we present a compact and formal description of VSA hardware accelerator's kernel functionality as follows:

{\footnotesize
\[ F\big(y,(s_1,s_2,s_3)\big):=
    \begin{cases} 
      a\big(y, (s_1, s_2)\big); & s_3=0 \\
      c(y); & s_3=1 \\
      e(y); & s_3=2
   \end{cases} \]
}

\begin{figure}[t!]
% \vspace{-1em}
	\centering
	\includegraphics[width=\columnwidth]{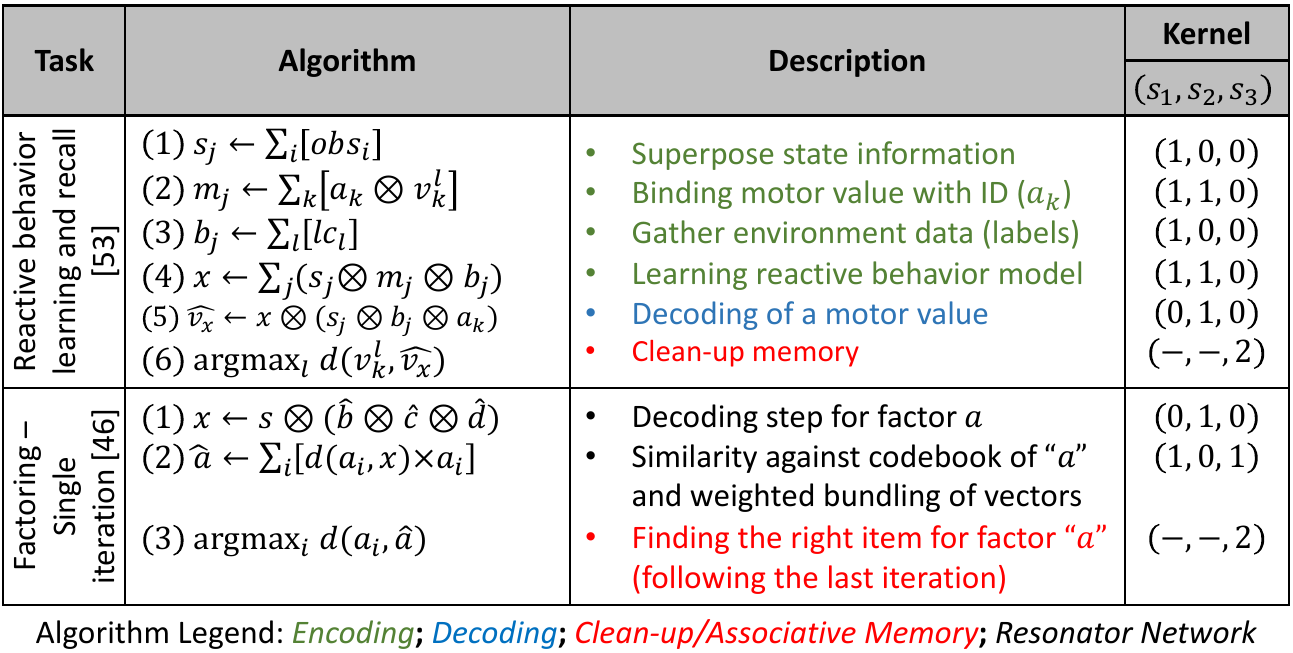}
	\caption{\textbf{Compact VSA Kernel Formulation.} Illustration of how the VSA kernel is programmed to implement workloads.}
 \vspace{-20pt}
	\label{fig:Kernel_Table}
\end{figure}

\noindent In this definition, the control variables $(s_1,s_2,s_3)$ are used to dynamically adjust the behavior of the kernel during runtime.
%based on the specific task being performed. 
Fig.~\ref{fig:Kernel_Table} demonstrates how the kernel is adjusted to execute VSA workloads. Performance results based on the mapping of these workloads and others are shown in Sec.~\ref{subsec:eval}.  

\vspace{-0.1in}
\subsection{Hardware Architecture and Dataflow}
\label{subsec:accelerator_dataflow}
\begin{figure}[!t]
	\centering
\includegraphics[width=0.9\columnwidth]{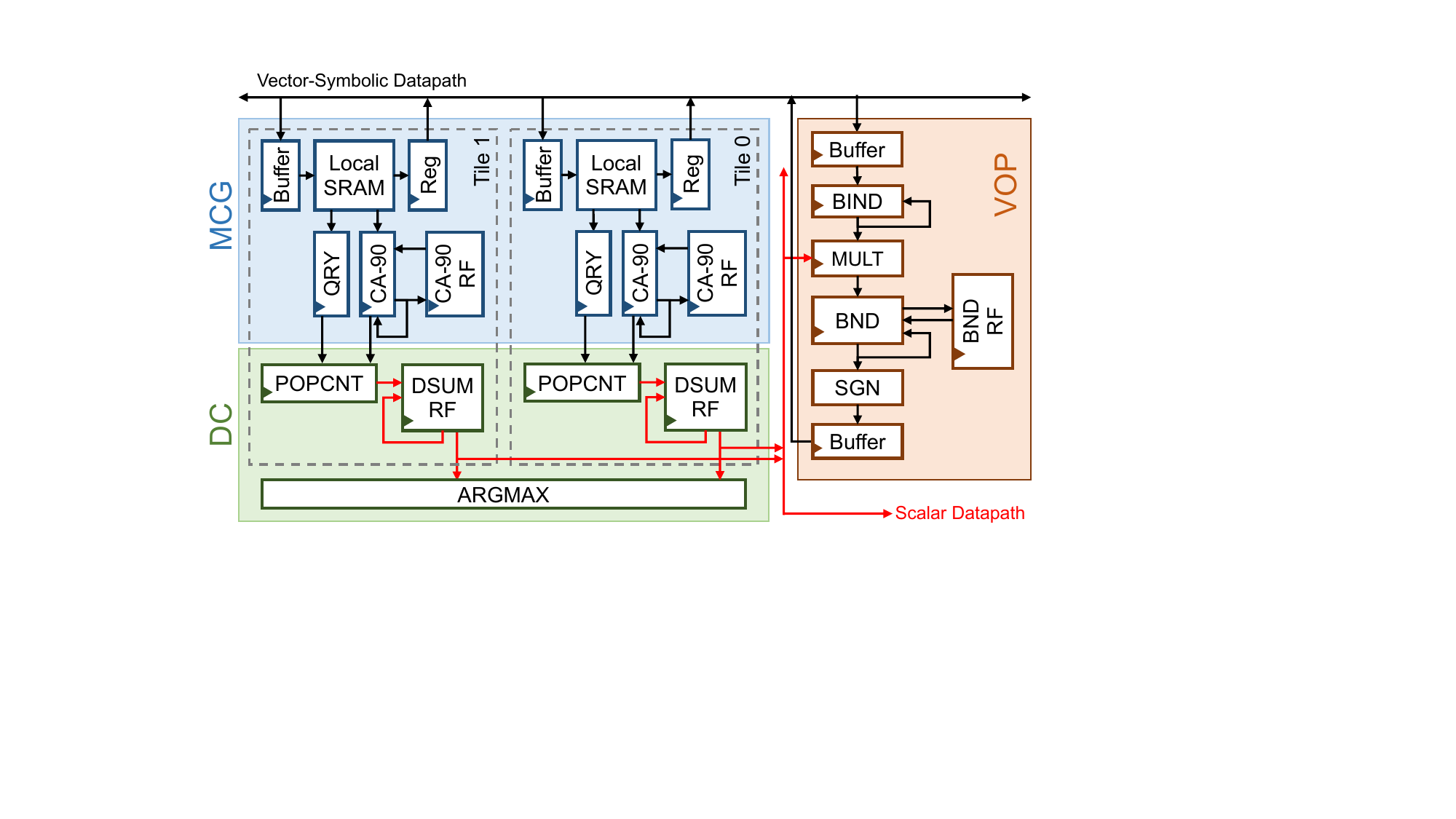}
	\caption{\textbf{Hardware Architecture and Dataflow.} The proposed multi-tile architecture for VSA consists of MCG, DC, and VOP subsystems.}
	\label{fig:hpu_arch}
 \vspace{-1em}
\end{figure}

We present a method for constructing architecture dataflow informed by the derived VSA kernels. Fig.~\ref{fig:hpu_arch} shows the overall architecture, consisting of three subsystems: (1) memory and codebook-generation subsystem (MCG), (2) vector-symbolic operations subsystem (VOP), and (3) distance computation subsystem (DC). A control unit is used to decode instructions and determine control configurations. A description of these subsystems and their internal operations are presented below.
%, (4) systolic array-based neural engine (SN), and (5) a global controller unit (CTRL). %A control unit is used to decode instructions and determine control configurations. 
%Fig.~\ref{fig:hpu_arch} depicts these subsystems and their internal modules. 

% \noindent
\textbf{MCG Subsystem.} This subsystem is distributed across multiple tiles, with each tile comprising four functional modules: a local memory (SRAM), a logic unit implementing cellular automata with rule 90 (CA-90)~\cite{kleyko2021cellular}, a register file (CA-90 RF), and a query register (QRY). Vectors loaded from the local SRAM are processed exclusively within the tile's logic to leverage near-memory computing. The SRAMs are initialized with randomly generated atomic vectors (i.e., codebooks) used for symbolic encoding. The dimension of these vectors is constrained by the size of the physical datapth; therefore, we utilize a folding mechanism to support extended vector dimensions. CA-90 is integral to this mechanism, utilizing XOR and shift operations to generate new random vectors on-the-fly~\cite{kleyko2021cellular}. This design significantly reduces the memory footprint, as only seed folds need to be stored in the local SRAMs. CA-90 RF is a register file that temporarily stores newly generated folds to minimize redundant activations of CA-90. The QRY register holds query data required for similarity computation, an essential component of VSA. 

% \noindent
\textbf{VOP Subsystem.} The VOP subsystem implements key VSA operations, used to construct distributed perceptual representations and perform symbolic reasoning computations. It consists of five logic units: a binding unit (BIND), a multiplying unit (MULT), a bundling unit (BND), a register file (BND RF), and a sign unit (SGN). BIND connects to a local buffer storing vectors and is to used to execute the binding operations over these vectors. The superposition of binded vectors is implemented in BND through element-wise addition (bundling). BIND and BND utilize different data representations, with BIND using binary and BND using integer formats. MULT manages the conversion from binary to integer formats and also performs element-wise scalar multiplication, an essential operation for neuro-symbolic encoding. Integer folds outputted from BND can be temporarily stored in BND RF for continuous superposition or converted to binary through SGN for transfer over the global vector-symbolic datapath. 

% \noindent
\textbf{DC Subsystem.} This subsystem handles operations for distance computation and nearest neighbor search, and it comprises three critical logic units: POPCNT, DSUM RF, and ARGMAX. POPCNT evaluates the popcount of the difference between two vectors; it executes element-wise XOR operations followed by an addition operation to compute the difference between the number of $1$'s and the number of $0$'s in the difference vector. As POPCNT operates on partial vectors due to vector folding, its output also represents a partial distance quantity. Hence, DSUM RF facilitates distance accumulation over multiple partial vectors, distributing distance computations across multiple independently controllable registers. The resulting distance data is then communicated to ARGMAX, which manages the search for the nearest neighbor vector based on the transferred distance values.

% \noindent
\textbf{Parameterized Multi-Tile Architecture.} 
The integration of the above modules leads to a ``single-tile'' architecture, which includes a single instance of MCG and DC. We also propose a ``multi-tile'' architecture, which allows memory-bounded vector loading and similarity computations to be distributed across multiple tiles and exploits a SIMD implementation to speed up the execution. This approach, therefore, enables parallel, near-memory processing of symbolic computations, and hence improves the utilization of compute units. A multi-tile architecture also extends the storage capabilities, providing a means to accommodate larger models. Tiles are also equipped with configuration registers, which allow tiles to be selectively activated (or deactivated) before issuing instructions.

\begin{figure}[!t]
	\centering
	\includegraphics[width=0.95\columnwidth]{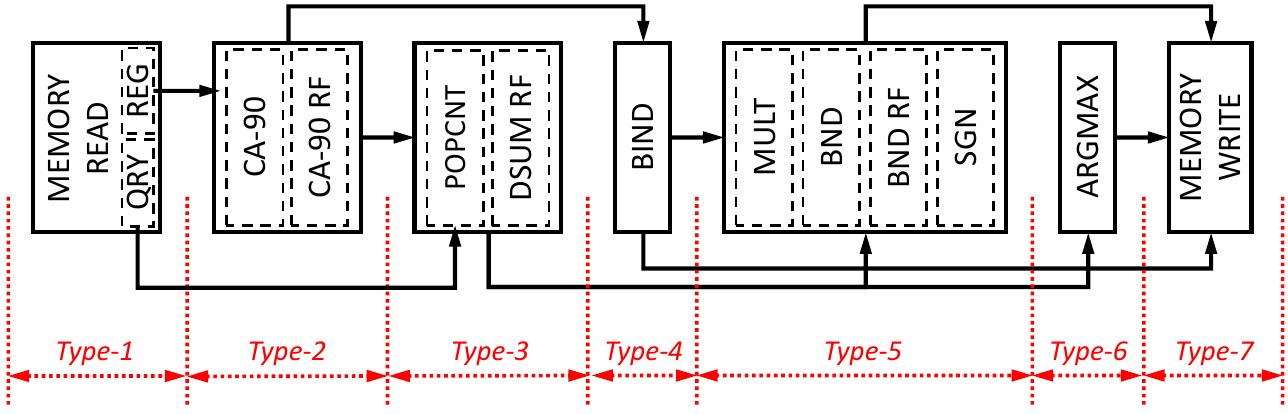}
	\caption{VSA accelerator's pipeline stages and operation types. 
 % {\blue{CK: I am thinking whether reviews will attack why we are having such a deep pipeline on a specific task, because generally this increases control hazard which in turns reduce IPC. So if we can explain this a bit, e.g. there's no hazard etc.,c., will be great}}
 }
	\label{fig:control_method}
 \vspace{-1em}
\end{figure}

\vspace{-0.1in}
\subsection{Accelerator Control Methods}
\label{subsec:accelerator_control}

The configuration of the different modules as described above exhibits a pipelined architecture that consists of seven pipeline stages, with each stage associated with a certain type of operation (Fig.~\ref{fig:control_method}). Such a pipelined configuration motivates a streamlined integration of dataflow and control-flow primitives, allowing different control methods to be applied without hazard.
% a pipelined configuration motivates the search for programming methods that improve resource utilization while equipping with control over performance tradeoffs. 
To perform this study, we particularly examine two control methods for this accelerator: \emph{single-operation-per-cycle} (SOPC) and \emph{multiple-operations-per-cycle} (MOPC).

\textbf{SOPC and MOPC.}
SOPC simplifies programming and reduces power consumption since only one pipeline stage switches during each cycle. However, this approach increases runtime, making it unsuitable for high-throughput applications. Conversely, MOPC enables pipeline stages to perform operations simultaneously, thus increasing the number of operations per cycle. However, MOPC leads to increased power consumption and requires a complex mapping framework to analyze program dependencies and optimize control activities. MOPC is better suited for high-throughput applications that require a balance between runtime and power consumption. 

\begin{figure}[!t]
	\centering
	\includegraphics[width=\columnwidth]{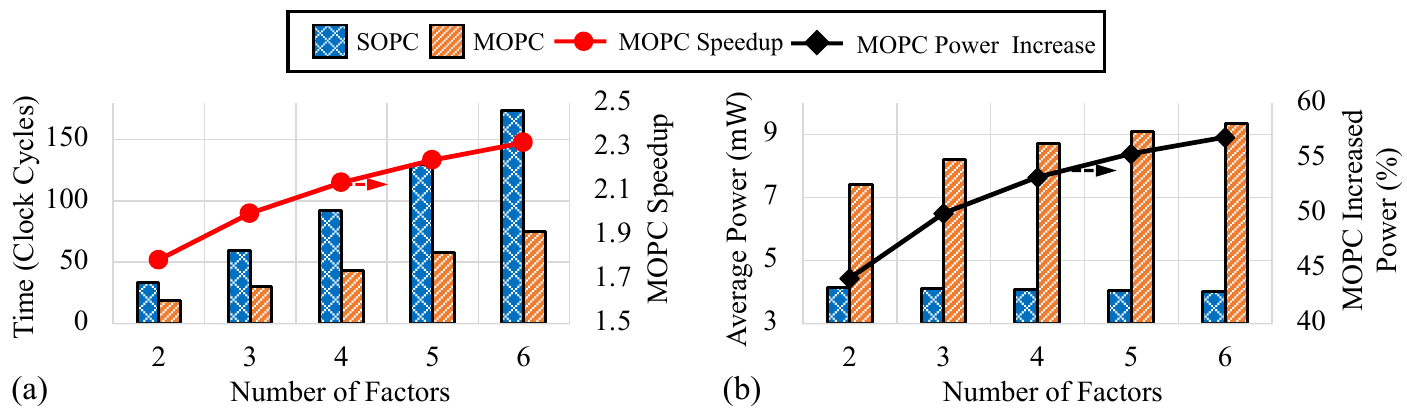}
	\caption{\textbf{Accelerator Control  Methods Comparison.} Runtime and power consumption results under two control methods (SOPC and MOPC) when executing the resonator network VSA algorithm.}
	\label{fig:SOPC_MOPC}
 \vspace{-5pt}
\end{figure}

% \noindent
\textbf{Control Methods Comparison.} We compare SOPC and MOPC by implementing factorization using the resonator network kernel. Fig.~\ref{fig:SOPC_MOPC} compares the runtime and power consumption of SOPC and MOPC when executing at various complexity levels (number of factors). We observe that MOPC achieves lower runtime in comparison with SOPC, and that the speed-up gained by using MOPC increases from $1.8$ to $2.3$. However, using MOPC also increases power consumption by $44\%$ to $57\%$ as the complexity increases. We adopt MOPC in our design because its better speedup capability is especially important when multiple heterogeneous tasks need to be executed simultaneously. Moreover, the speed-up gain of MOPC can be flexibly configured based on power-consumption constraints for low-power purposes.

% \noindent
% \textbf{Adopting MOPC.} We optimize the control-flow of our design based on MOPC as it offers two major advantages. First, MOPC provides the better capability to speed up execution. This is especially important when multiple heterogeneous tasks need to be executed simultaneously. Second, the speed-up gain can be flexibly configured based on power-consumption constraints. Consider the case of executing a six-factor resonator network using MOPC. If the speed-up is decreased from $2.3$ to $2.0$, the average power consumption decreases from $9.3$ mW to $8.0$ mW. In fact, we could aim to decrease the speed-up further until it converges to that of SOPC; that is $1.0$, achieving a reduced power consumption of only $4.0$ mW. In other words, SOPC is considered a special case of MOPC. 

\begin{figure}[t!]
 % \vspace{-1em}
	\centering
\includegraphics[width=0.9\columnwidth]{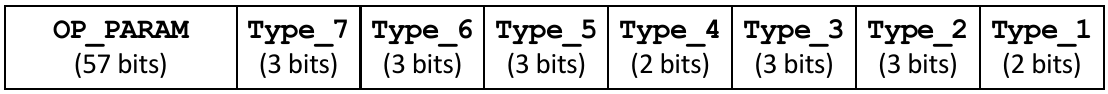}
	\caption{The \emph{Instruction Word} format adopted in the proposed design. }
	\label{fig:MPOC_ISA}
 \vspace{-10pt}
\end{figure}

% \subsection{Accelerator Instruction Format}
% \noindent
\textbf{Accelerator Instruction Format.} 
To realize the MOPC control method, we design an instruction-set architecture that employs a wide-word macro format, referred to as \emph{Instruction Word}. Similar to a Very-Large Instruction Word (VLIW), a single \emph{Word} consists of multiple operations, except that these operations are sequential in the pipelined dataflow and not parallel like VLIW architectures. As shown in Fig.~\ref{fig:MPOC_ISA}, the \emph{Word} format consists of seven \texttt{Type} fields, used to specify the operations to be executed in seven pipelined stages, and an \texttt{OP\_PARAM} field, used to configure \texttt{Type} operations. This approach offers a high degree of flexibility and is commonly used with domain-specific processors. Details on the instruction fields and compiler optimization are omitted due to space. 

\vspace{-0.1in}
\subsection{Evaluation Results}
\label{subsec:eval}
\textbf{Experimental Setup.} The design was implemented in SystemVerilog and synthesized with Synopsys Design Compiler using foundry 28nm library. Tab.~\ref{tab:eval_config} lists the architectural parameters. 
% A sensitivity analysis was performed to explore the impact of the processor's integer bit-width on the accuracy and select the proper values of \textit{\textbf{H}} and \textit{\textbf{C}}. 
The energy is measured using Synopsys PrimeTime PX. 
VSA workloads are also simulated on NVIDIA V100 GPU as the baseline, and GPU power was measured using the \texttt{nvidia-smi} utility. The algorithms listed in Tab.~\ref{tab:eval_workload} are used for evaluation, facilitating a comprehensive assessment of multi-layer cognition systems.

% \begin{figure}[t!]
% \vspace{-1em}
% 	\centering
% 	\includegraphics[width=\columnwidth]{Figs/Benchmarks.pdf}
% 	\caption{Evaluation setup: (a) synthesized HPU configurations; (b) HD workloads used in simulation.}
% 	\label{fig:Benchmarks}
% \end{figure}

\begin{figure}[!t]
 \vspace{-1em}
	\centering
	\includegraphics[width=.95\columnwidth]{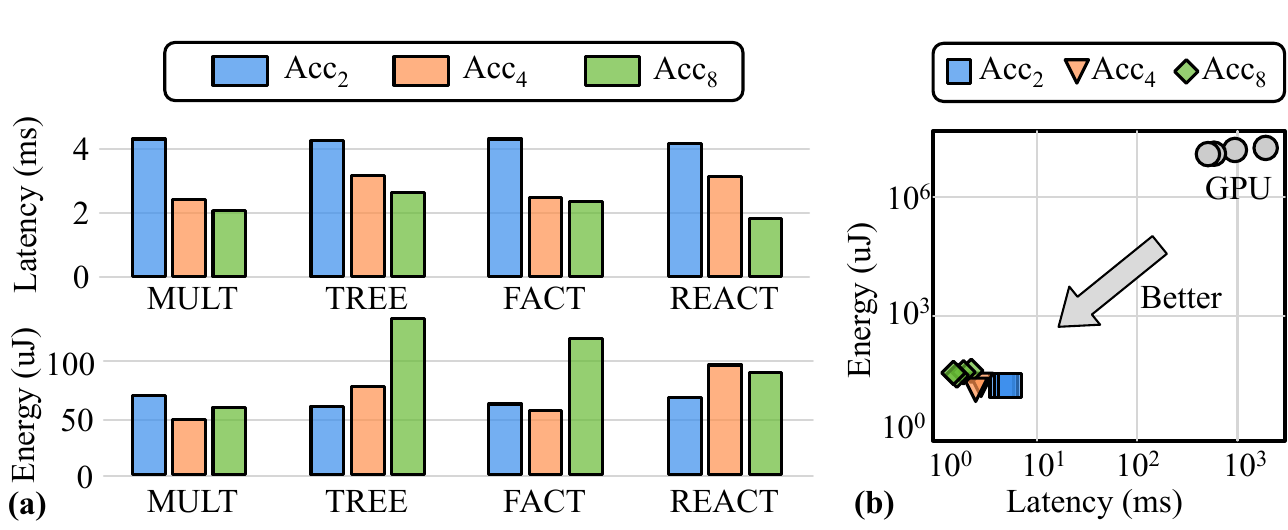}
	\caption{\textbf{VSA Accelerator (Acc) Efficiency.} \textbf{(a)} Comparison between Acc$_2$, Acc$_4$, and Acc$_8$ in terms of latency and energy consumption across workloads. \textbf{(b)} Comparison between Acc and GPU (baseline).}
	\label{fig:acc_results}
 \vspace{-1em}
\end{figure}

% \noindent
\textbf{Latency.}
We first evaluate the impact of varying VSA accelerator (Acc) size on latency. Fig.~\ref{fig:acc_results}\blue{a} shows that Acc$_4$ provides speed-up of 1.3-1.8$\times$ compared to Acc$_2$, highlighting resource underprovisioning in Acc$_2$. However, we observe that the benefits of scaling up the design from Acc$_4$ to Acc$_8$ are not equally realized by all algorithms. Specifically, only $1.16\times$ speed-up is achieved by MULT. This is because MULT typically performs VOP-intensive computations for sequence encoding and thus its response to further increase in design size is minimal. On the other hand, REACT achieves $1.69\times$ speed-up when Acc$_8$ is used. This result is attributed to the fact that REACT performs extensive clean-up memory operations, which can be efficiently distributed across all tiles.

% \noindent
\textbf{Energy Consumption.}
Fig.~\ref{fig:acc_results}\blue{a} shows that the energy efficiency does not exhibit systematic behavior as the accelerator size varies. The reasons are twofold: (1) The leakage power becomes increasingly significant when Acc size is increased. Our analysis shows that the leakage power increases from $1.7$~mW to $5.2$~mW (i.e., $3\times$ increase) when the design is scaled up from Acc$_2$ to Acc$_8$. (2) Each of the instructions has a unique effect on energy consumption, especially because instructions trigger circuit activity at different hardware modules.%; Fig.~\ref{}(b) shows the variation of energy per clock cycle across HPU modules. 

% These results underscore the critical role of compiler optimization in improving the energy efficiency of HPU. Specifically, they suggest that achieving maximum energy efficiency requires a unique optimization approach, which characterizes energy at both the module and system levels and employs targeted optimizations during compilation.

\begin{table}[t!]
% \scriptsize
\centering
\caption{\textbf{Hardware Setup.} VSA accelerator (Acc) configurations.}
% \vspace{0.1in}
\renewcommand*{\arraystretch}{1.05}
\setlength\tabcolsep{1.8pt}
\resizebox{\linewidth}{!}{%
\begin{tabular}{c|c|c|c|c|c|c|c|c}
\hline
\begin{tabular}[c]{@{}c@{}}\textbf{Instance}   \\      \textbf{Name}\end{tabular} & \begin{tabular}[c]{@{}c@{}}Bus Width \\  (\textit{\textbf{W}})\end{tabular} & \begin{tabular}[c]{@{}c@{}}$\#$Tiles \\      (\textit{\textbf{K}})\end{tabular} & \begin{tabular}[c]{@{}c@{}}$\#$CA-90-RF \\  register (\textit{\textbf{R}})\end{tabular} & \begin{tabular}[c]{@{}c@{}}$\#$BND-RF \\ register (\textit{\textbf{B}}) \end{tabular} & \begin{tabular}[c]{@{}c@{}}$\#$DSUM \\ register (\textit{\textbf{D}}) \end{tabular} & \begin{tabular}[c]{@{}c@{}}Distance bit \\     -width (\textit{\textbf{C}})\end{tabular} & \begin{tabular}[c]{@{}c@{}}BND bit-\\ width (\textit{\textbf{H}})\end{tabular} & \begin{tabular}[c]{@{}c@{}} Memory \\      Capacity\end{tabular} \\ \hline
\textbf{Acc}$_{2}$ & 512 & 2 & 2 & 2 & 2 & 12 & 8 & 128~KB \\\hline
\textbf{Acc}$_{4}$ & 512 & 4 & 4 & 4 & 4 & 12 & 8 & 256~KB \\\hline
\textbf{Acc}$_{8}$ & 512 & 8 & 8 & 8 & 8 & 12 & 8 & 512~KB\\ \hline                                                     
\end{tabular}
}
\label{tab:eval_config}
% \vspace{-10pt}
\end{table}

\begin{table}[t!]
\scriptsize
\centering
\caption{\textbf{Algorithm Setup}. VSA workloads used in evaluation.}
% \vspace{0.1in}
\renewcommand*{\arraystretch}{1.05}
\setlength\tabcolsep{2pt}
\resizebox{\linewidth}{!}{%
\begin{tabular}{c|c|c|c}
\hline
\textbf{Workload} & \textbf{Layer}  & \textbf{Application}                                  & \textbf{Problem Size (Complexity)}                                  \\ \hline
\textbf{MULT}     & Perception                  & \begin{tabular}[c]{@{}c@{}}Multi-modal learning and\\ Inference~\cite{datta2019programmable}\end{tabular} & \begin{tabular}[c]{@{}c@{}}300 samples, 120 item vectors, 16\\  prototype vectors (classes), 100 queries \end{tabular} \\ \hline
\textbf{TREE}     & \multirow{3}{*}{Reasoning} & Tree encoding and search~\cite{kleyko2021vector}                                                  &   70 tree structures, 9 items, 400 queries                                                \\ \cline{1-1} \cline{3-4} 
\textbf{FACT}     &                    & Factorization of data sets~\cite{frady2020resonator}                                                  & \begin{tabular}[c]{@{}c@{}}60 iterations, 120 item vectors, 13\\      prototype vectors\end{tabular} \\ \hline
\textbf{REACT}     & Control                  & Motor learning and recall~\cite{menon2023shared}                                                  & 500 samples, 55 item vectors, 160 recalls                                                \\ \hline
\end{tabular}
}
\label{tab:eval_workload}
% \vspace{-10pt}
\end{table}

% \noindent
\textbf{Comparison with GPU.}
We also compare all VSA accelerator instances with GPU in terms of latency and energy consumption. Fig.~\ref{fig:acc_results}\blue{b} shows that Acc is up to three orders of magnitude faster in executing VSA workloads than GPU, despite using batch processing in our GPU implementation. This result consolidates suggests the GPU-memory interface is not optimized for VSA data transfer. In addition, Acc operation is up to six orders of magnitude more energy efficient than GPU processing.
% These findings also align with comparisons reported in prior classification processors~\cite{datta2019programmable}.
%work~\cite{datta2019programmable}. 
This performance gap is attributed to GPU's scalar architecture, which relies on complex SIMD arithmetic units to perform simple vector operations.

% \input{Sections/6_Hardware}
% \vspace{-0.5em}
\section{Outlook and Research Opportunities}
% \vspace{-0.3em}
\label{sec:challenge}
% In this section, we discuss the challenges and opportunities for NSAI systems, and our view of the road ahead. We focus our discussion from system and architecture perspectives.
In this section, we discuss the challenges and opportunities for neuro-symbolic systems, and outline our vision for the future, focusing on the system and architecture perspectives.

% dataset
% \textbf{Building dataset for the cognitive capability of NSAI.}
% NSAI system holds great potential in achieving human-like AI~\cite{booch2021thinking}. However, their application is currently limited to a few tasks that focus mainly on basic decision-making and reasoning problems~\cite{garcez2022neural}. This falls short of the broader vision of human cognitive abilities, such as interpretability, deductive reasoning, systematicity, productivity, compositionality, inferential coherence of mental thought, causal and counterfactual thinking. We need more challenging and suitable datasets and testbeds with data-centric approaches to significantly advance the metacognitive capabilities of NSAI systems.
\textbf{Building ImageNet-like neuro-symbolic datasets.}
Neuro-symbolic systems hold great potential in achieving human-like performance~\cite{booch2021thinking}. However, their current applications are still limited to basic decision-making and reasoning problems, falling short of the broader vision of human cognitive abilities, such as deductive reasoning, compositionality, and counterfactual thinking. It is still an open question of how perception learned from other domains can be transferred to abstract reasoning tasks. To significantly advance the metacognitive capabilities of neuro-symbolic systems, more challenging and suitable datasets are highly desirable to unleash its potential.

% algorithm
% \textbf{Unifying neuro-symbolic-probabilistic models.}
% Integrating neural, symbolic, and probabilistic approaches offers promise to improve the explainability and robustness of ML models. However, current efforts to combine these complementary approaches are still in a nascent manner~\cite{wang2022towards} - how to integrate them in a principled manner is a fundamental and open challenge. We envision a unified framework to design algorithms that opportunistically combine neural, symbolic, and probabilistic components, and quantifying scaling laws for neuro-probabilistic-symbolic inference versus large neural models. 
\textbf{Unifying neuro-symbolic models.}
Integrating neural, symbolic, and probabilistic approaches offers promise to improve AI models' explainability and robustness. However, the current attempts to combine these complementary approaches are still in a nascent manner - how to integrate them in a principled manner remains an open challenge. Particularly, symbolic components can be combined with Large Language Models (LLMs) to improve their planning and reasoning capabilities~\cite{kambhampati2024llms}.
We envision a unified framework to design algorithms that opportunistically combine neural and symbolic with probabilistic representations, and for quantifying scaling laws for neuro-symbolic inference versus large neural models. 

% software tool
% \textbf{Developing efficient software frameworks.}
% NSAI systems typically utilize underlying logic, such as fuzzy logic, parameterization, and differentiable structures, to support learning and reasoning capabilities. 
% However, most NSAI system implementations create custom software for deduction for the particular logic used, which limits modularity and extensibility~\cite{aditya2023pyreason}. 
% Therefore, we need new software frameworks that can encompass a broad set of reasoning logical capabilities, and provide practical syntactic and semantic extensions, while being fast and memory-efficient. Moreover, new programming models, compilers, and runtimes that can enable the easy and efficient realization of the neuro-symbolic-probabilistic models are of significance to realize the full promise of NSAI paradigms.
\textbf{Developing efficient software frameworks.}
Neuro-symbolic AI systems typically utilize underlying logic, such as fuzzy logic, parameterization, and differentiable structures, to support learning and reasoning capabilities. 
However, most system implementations create custom software for deduction for the particular logic, which limits modularity and extensibility. 
Thus, new software frameworks are needed that can encompass a broad set of reasoning logical capabilities and provide practical syntactic and semantic extensions while being fast and memory-efficient. Moreover, new programming models and compilers that can facilitate the ease and efficient realization of the neuro-symbolic models are of significance to realize the full promise of neuro-symbolic AI paradigms.

% benchmarking
% \textbf{Benchmarking diverse NSAI workloads.} 
% Given the proliferation of NSAI algorithms and the rapid advances of hardware platforms, benchmarking NSAI systems in a comparable, quantitative, and validatable manner is imperative. 
% From the system aspect, we need a set of representative benchmarks that capture the essential workload characteristics (compute kernels, access patterns, sparsity) of neural, symbolic, and probabilistic models, and quantitatively test in human-AI applications.
% From the architecture and hardware aspect, we need modeling-simulation frameworks to enable novel architecture for the neural, symbolic, and probabilistic workloads, and build optimized modular blocks as libraries by leveraging workload characteristics. 
% Benchmarking NSAI computing will guide ML researchers and system architects to investigate the trade-offs in accuracy, performance, and efficiency of various NSAI algorithms and implement the system in a performance-portable way.
\textbf{Benchmarking diverse neuro-symbolic workloads.} 
Given the proliferation of neuro-symbolic algorithms and the rapid hardware advancements, it is crucial to benchmark neuro-symbolic AI systems in a comparable and validated manner.
To achieve this, from the system aspect, we need representative benchmarks that capture the essential workload characteristics (e.g., compute kernels, access patterns, and sparsity) of neural and symbolic models, and that can be quantitatively tested in human-AI applications.
From an architectural and hardware perspective, we need modeling-simulation frameworks to enable the development of novel architectures for these workloads and build optimized modular blocks as libraries by leveraging workload characteristics.
Benchmarking neuro-symbolic computing will guide ML researchers and system architects in investigating the trade-offs in accuracy, performance, and efficiency of various neuro-symbolic algorithms, and in implementing systems in a performance-portable way.

% 
% Such benchmarking comes into two folds, workloads and architecture.
% will greatly ease the design process

% architecture
% \textbf{Designing cognitive hardware architecture.}
% NSAI workloads that combine neural, symbolic, and probabilistic methods will feature much greater heterogeneity in compute kernels, sparsity, irregularity in access patterns, and higher memory intensity than current DNN workloads. This will lead to an increasing divergence with the current hardware roadmap (e.g., systolic arrays or compute-in-memory crossbars) which largely focuses on matrix multiplication/nearest neighbor search and regular dataflows. Therefore, we need novel architectures with processing units with more features, flexibility, memory hierarchies, and on-chip interconnects for handling these additional complexities in computing and communication. Besides, the hardware architecture needs to provide flexibility in the form of configurable interconnects and full addressable memories to allow for NSAI algorithmic innovations without making hardware obsolete.
\textbf{Designing cognitive hardware architectures.}
Neuro-symbolic workloads that combine neural, symbolic, and probabilistic methods feature much greater heterogeneity in compute kernels, sparsity, irregularity in access patterns, and higher memory intensity than DNNs. This leads to an increasing divergence with the current hardware roadmap that largely focuses on matrix multiplication and regular dataflow. Therefore, we need novel architectures with dedicated processing units, memory hierarchies, and NoCs that can handle the additional complexities in computations and communications. Additionally, the architecture needs to provide flexibility with both configurable interconnects and full addressable memories to keep pace with neuro-symbolic AI algorithmic innovations.

\vspace{-0.05in}
\section{Conclusion}
% \vspace{-0.3em}
\label{sec:conclusion}
% Neuro-symbolic AI is an emerging computing paradigm for next-generation efficient, robust, explainable, and cognitive AI systems. This paper systematically reviews the recent NSAI algorithms, characterizes the system performance, analyzes workload operators, and identifies the challenges and opportunities towards next-generation NSAI systems.
% Neuro-symbolic AI is an emerging paradigm for next-generation efficient, robust, explainable, and cognitive AI systems. This paper systematically characterizes neuro-symbolic system performance and analyzes their workload operators. Built upon profiling insights, we suggest cross-layer optimization techniques and present a hardware architecture case study for their performance and efficiency. We anticipate this research to unlock the challenges and opportunities towards fulfilling next-generation neuro-symbolic AI systems.
Neuro-symbolic AI is an emerging paradigm for developing efficient, robust, explainable, and cognitively advanced AI systems. This paper provides a systematic characterization of neuro-symbolic system performance and analyzes their operational components. Leveraging insights from profiling, we propose cross-layer optimization techniques and present a case study of a hardware architecture designed to enhance their performance and efficiency. We believe this research will address key challenges and highlight opportunities essential for advancing next-generation neuro-symbolic AI systems.

% \clearpage
\section*{Acknowledgements}
This work was supported in part by CoCoSys, one of seven centers in JUMP 2.0, a Semiconductor Research Corporation (SRC) program sponsored by DARPA.

\ifCLASSOPTIONcaptionsoff
  \newpage
\fi



%\bibliographystyle{IEEEtran}


% \end{thebibliography}
\bibliographystyle{ieeetr}
\bibliography{refs}

\begin{thebibliography}{10}

\bibitem{wu2022sustainable}
C.-J. Wu, R.~Raghavendra, U.~Gupta, B.~Acun, N.~Ardalani, K.~Maeng, G.~Chang, F.~Aga, J.~Huang, C.~Bai, {\em et~al.}, ``Sustainable ai: Environmental implications, challenges and opportunities,'' {\em Proceedings of Machine Learning and Systems (MLSys)}, vol.~4, pp.~795--813, 2022.

\bibitem{wan2021analyzing}
Z.~Wan, A.~Anwar, Y.-S. Hsiao, T.~Jia, V.~J. Reddi, and A.~Raychowdhury, ``Analyzing and improving fault tolerance of learning-based navigation systems,'' in {\em 2021 58th ACM/IEEE Design Automation Conference (DAC)}, pp.~841--846, IEEE, 2021.

\bibitem{debenedetti2023scaling}
E.~Debenedetti, Z.~Wan, M.~Andriushchenko, V.~Sehwag, K.~Bhardwaj, and B.~Kailkhura, ``Scaling compute is not all you need for adversarial robustness,'' {\em arXiv preprint arXiv:2312.13131}, 2023.

\bibitem{wan2024towards_2}
Z.~Wan, C.-K. Liu, R.~Raj, C.~Li, H.~You, Y.~Fu, C.~Wan, A.~Samajdar, C.~Lin, T.~Krishna, and A.~Raychowdhury, ``Towards cognitive ai systems: Workload and characterization of neuro-symbolic ai,'' in {\em Proceedings of IEEE International Symposium on Performance Analysis of Systems and Software (ISPASS)}, 2024.

\bibitem{garcez2023neurosymbolic}
A.~d. Garcez and L.~C. Lamb, ``Neurosymbolic ai: The 3 rd wave,'' {\em Artificial Intelligence Review}, pp.~1--20, 2023.

\bibitem{wan2024towards}
Z.~Wan, C.-K. Liu, H.~Yang, C.~Li, H.~You, Y.~Fu, C.~Wan, T.~Krishna, Y.~Lin, and A.~Raychowdhury, ``Towards cognitive ai systems: a survey and prospective on neuro-symbolic ai,'' {\em arXiv preprint arXiv:2401.01040}, 2024.

\bibitem{hersche2023neuro}
M.~Hersche, M.~Zeqiri, L.~Benini, A.~Sebastian, and A.~Rahimi {\em Nature Machine Intelligence}, pp.~1--13, 2023.

\bibitem{maoneuro}
J.~Mao, C.~Gan, P.~Kohli, J.~B. Tenenbaum, and J.~Wu, ``The neuro-symbolic concept learner: Interpreting scenes, words, and sentences from natural supervision,'' in {\em International Conference on Learning Representations (ICLR)}, 2019.

\bibitem{gu2023conceptgraphs}
Q.~Gu, A.~Kuwajerwala, S.~Morin, K.~M. Jatavallabhula, B.~Sen, A.~Agarwal, C.~Rivera, W.~Paul, K.~Ellis, R.~Chellappa, {\em et~al.}, ``Conceptgraphs: Open-vocabulary 3d scene graphs for perception and planning,'' {\em arXiv preprint arXiv:2309.16650}, 2023.

\bibitem{samajdar2020systematic}
A.~Samajdar, J.~M. Joseph, Y.~Zhu, P.~Whatmough, M.~Mattina, and T.~Krishna, ``A systematic methodology for characterizing scalability of dnn accelerators using scale-sim,'' in {\em 2020 IEEE International Symposium on Performance Analysis of Systems and Software (ISPASS)}, pp.~58--68, IEEE, 2020.

\bibitem{kwon2021heterogeneous}
H.~Kwon, L.~Lai, M.~Pellauer, T.~Krishna, Y.-H. Chen, and V.~Chandra, ``Heterogeneous dataflow accelerators for multi-dnn workloads,'' in {\em 2021 IEEE International Symposium on High-Performance Computer Architecture (HPCA)}, pp.~71--83, IEEE, 2021.

\bibitem{wu2023highlight}
Y.~N. Wu, P.-A. Tsai, S.~Muralidharan, A.~Parashar, V.~Sze, and J.~Emer, ``Highlight: Efficient and flexible dnn acceleration with hierarchical structured sparsity,'' in {\em Proceedings of the 56th Annual IEEE/ACM International Symposium on Microarchitecture (MICRO)}, pp.~1106--1120, 2023.

\bibitem{ramachandran2024algorithm}
A.~Ramachandran, Z.~Wan, G.~Jeong, J.~Gustafson, and T.~Krishna, ``Algorithm-hardware co-design of distribution-aware logarithmic-posit encodings for efficient dnn inference,'' {\em arXiv preprint arXiv:2403.05465}, 2024.

\bibitem{fan2024benchmarking}
Z.~Fan, Z.~Wan, C.-K. Liu, A.~Lu, K.~Bhardwaj, and A.~Raychowdhury, ``Benchmarking test-time dnn adaptation at edge with compute-in-memory,'' {\em Journal on Autonomous Transportation Systems}, 2024.

\bibitem{ibrahim2024efficient}
M.~Ibrahim, Y.~Kim, and J.~M. Rabaey, ``Efficient design of a hyperdimensional processing unit for multi-layer cognition,'' in {\em 2024 Design, Automation \& Test in Europe Conference \& Exhibition (DATE)}, pp.~1--6, IEEE, 2024.

\bibitem{silver2017mastering}
D.~Silver, T.~Hubert, J.~Schrittwieser, I.~Antonoglou, M.~Lai, A.~Guez, M.~Lanctot, L.~Sifre, D.~Kumaran, T.~Graepel, {\em et~al.}, ``Mastering chess and shogi by self-play with a general reinforcement learning algorithm,'' {\em arXiv preprint arXiv:1712.01815}, 2017.

\bibitem{pryor2022neupsl}
C.~Pryor, C.~Dickens, E.~Augustine, A.~Albalak, W.~Wang, and L.~Getoor, ``Neupsl: Neural probabilistic soft logic,'' {\em arXiv preprint arXiv:2205.14268}, 2022.

\bibitem{yang2020neurasp}
Z.~Yang, A.~Ishay, and J.~Lee, ``Neurasp: Embracing neural networks into answer set programming,'' in {\em 29th International Joint Conference on Artificial Intelligence (IJCAI)}, 2020.

\bibitem{dai2019bridging}
W.-Z. Dai, Q.~Xu, Y.~Yu, and Z.-H. Zhou, ``Bridging machine learning and logical reasoning by abductive learning,'' {\em Advances in Neural Information Processing Systems (NeurIPS)}, vol.~32, 2019.

\bibitem{yi2018neural}
K.~Yi, J.~Wu, C.~Gan, A.~Torralba, P.~Kohli, and J.~Tenenbaum, ``Neural-symbolic vqa: Disentangling reasoning from vision and language understanding,'' {\em Advances in Neural Information Processing Systems (NeurIPS)}, vol.~31, 2018.

\bibitem{theiss2022unpaired}
J.~Theiss, J.~Leverett, D.~Kim, and A.~Prakash, ``Unpaired image translation via vector symbolic architectures,'' in {\em European Conference on Computer Vision (ECCV)}, pp.~17--32, Springer, 2022.

\bibitem{zhang2021abstract}
C.~Zhang, B.~Jia, S.-C. Zhu, and Y.~Zhu, ``Abstract spatial-temporal reasoning via probabilistic abduction and execution,'' in {\em Proceedings of the IEEE/CVF Conference on Computer Vision and Pattern Recognition (CVPR)}, pp.~9736--9746, 2021.

\bibitem{riegel2020logical}
R.~Riegel, A.~Gray, F.~Luus, N.~Khan, N.~Makondo, I.~Y. Akhalwaya, H.~Qian, R.~Fagin, F.~Barahona, U.~Sharma, {\em et~al.}, ``Logical neural networks,'' {\em arXiv preprint arXiv:2006.13155}, 2020.

\bibitem{lampledeep}
G.~Lample and F.~Charton, ``Deep learning for symbolic mathematics,'' in {\em International Conference on Learning Representations (ICLR)}, 2019.

\bibitem{evans2018learning}
R.~Evans and E.~Grefenstette, ``Learning explanatory rules from noisy data,'' {\em Journal of Artificial Intelligence Research}, vol.~61, pp.~1--64, 2018.

\bibitem{badreddine2022logic}
S.~Badreddine, A.~d. Garcez, L.~Serafini, and M.~Spranger, ``Logic tensor networks,'' {\em Artificial Intelligence}, vol.~303, p.~103649, 2022.

\bibitem{hohenecker2020ontology}
P.~Hohenecker and T.~Lukas, ``Ontology reasoning with deep neural networks,'' {\em Journal of Artificial Intelligence Research}, vol.~68, pp.~503--540, 2020.

\bibitem{lamb2020graph}
L.~C. Lamb, A.~Garcez, M.~Gori, M.~Prates, P.~Avelar, and M.~Vardi, ``Graph neural networks meet neural-symbolic computing: A survey and perspective,'' in {\em IJCAI 2020-29th International Joint Conference on Artificial Intelligence}, 2020.

\bibitem{wu2022zeroc}
T.~Wu, M.~Tjandrasuwita, Z.~Wu, X.~Yang, K.~Liu, R.~Sosic, and J.~Leskovec, ``Zeroc: A neuro-symbolic model for zero-shot concept recognition and acquisition at inference time,'' {\em Advances in Neural Information Processing Systems (NeurIPS)}, vol.~35, 2022.

\bibitem{dongneural}
H.~Dong, J.~Mao, T.~Lin, C.~Wang, L.~Li, and D.~Zhou, ``Neural logic machines,'' in {\em International Conference on Learning Representations (ICLR)}, 2019.

\bibitem{henry2020taxonomy}
H.~Kaut, ``Robert s. engelmore memorial lecture at aaai 2020,'' {\em https://roc-hci.com/announcements/the-third-ai-summer/}, 2020.

\bibitem{zhang2020alphazero}
H.~Zhang and T.~Yu, ``Alphazero,'' {\em Deep Reinforcement Learning: Fundamentals, Research and Applications}, pp.~391--415, 2020.

\bibitem{zhang2019raven}
C.~Zhang, F.~Gao, B.~Jia, Y.~Zhu, and S.-C. Zhu, ``Raven: A dataset for relational and analogical visual reasoning,'' in {\em Proceedings of the IEEE/CVF Conference on Computer Vision and Pattern Recognition (CVPR)}, pp.~5317--5327, 2019.

\bibitem{hu2021stratified}
S.~Hu, Y.~Ma, X.~Liu, Y.~Wei, and S.~Bai, ``Stratified rule-aware network for abstract visual reasoning,'' in {\em Proceedings of the AAAI Conference on Artificial Intelligence (AAAI)}, vol.~35, pp.~1567--1574, 2021.

\bibitem{manhaeve2021neural}
R.~Manhaeve, S.~Duman{\v{c}}i{\'c}, A.~Kimmig, T.~Demeester, and L.~De~Raedt, ``Neural probabilistic logic programming in deepproblog,'' {\em Artificial Intelligence}, vol.~298, p.~103504, 2021.

\bibitem{yi2020clevrer}
K.~Yi, C.~Gan, Y.~Li, P.~Kohli, J.~Wu, A.~Torralba, and J.~B. Tenenbaum, ``Clevrer: Collision events for video representation and reasoning,'' in {\em International Conference on Learning Representations (ICLR)}, 2020.

\bibitem{garcez2019neural}
A.~d. Garcez, M.~Gori, L.~C. Lamb, L.~Serafini, M.~Spranger, and S.~N. Tran, ``Neural-symbolic computing: An effective methodology for principled integration of machine learning and reasoning,'' {\em arXiv preprint arXiv:1905.06088}, 2019.

\bibitem{guo2005lubm}
Y.~Guo, Z.~Pan, and J.~Heflin, ``Lubm: A benchmark for owl knowledge base systems,'' {\em Journal of Web Semantics}, vol.~3, no.~2-3, 2005.

\bibitem{sutcliffe2017tptp}
G.~Sutcliffe, ``The tptp problem library and associated infrastructure,'' {\em Journal of Automated Reasoning}, vol.~59, no.~4, pp.~483--502, 2017.

\bibitem{asuncion2007uci}
A.~Asuncion and D.~Newman, ``Uci machine learning repository,'' 2007.

\bibitem{gencer2023research}
{\"O}.~GENCER, ``The research about morphometric characteristics on leptograpsus crabs,'' 2023.

\bibitem{manhaeve2018deepproblog}
R.~Manhaeve, S.~Dumancic, A.~Kimmig, T.~Demeester, and L.~De~Raedt, ``Deepproblog: Neural probabilistic logic programming,'' {\em Advances in neural information processing systems (NeurIPS)}, vol.~31, 2018.

\bibitem{pgm}
D.~Barrett, F.~Hill, A.~Santoro, A.~Morcos, and T.~Lillicrap, ``Measuring abstract reasoning in neural networks,'' in {\em International conference on machine learning (ICML)}, pp.~511--520, PMLR, 2018.

\bibitem{graves2016hybrid}
A.~Graves, G.~Wayne, M.~Reynolds, T.~Harley, I.~Danihelka, A.~Grabska-Barwi{\'n}ska, S.~G. Colmenarejo, E.~Grefenstette, T.~Ramalho, J.~Agapiou, {\em et~al.}, ``Hybrid computing using a neural network with dynamic external memory,'' {\em Nature}, vol.~538, no.~7626, pp.~471--476, 2016.

\bibitem{richter2016playing}
S.~R. Richter, V.~Vineet, S.~Roth, and V.~Koltun, ``Playing for data: Ground truth from computer games,'' in {\em European Conference on Computer Vision (ECCV)}, pp.~102--118, Springer, 2016.

\bibitem{cordts2016cityscapes}
M.~Cordts, M.~Omran, S.~Ramos, T.~Rehfeld, M.~Enzweiler, R.~Benenson, U.~Franke, S.~Roth, and B.~Schiele, ``The cityscapes dataset for semantic urban scene understanding,'' in {\em Proceedings of the IEEE conference on computer vision and pattern recognition (CVPR)}, 2016.

\bibitem{isola2017image}
P.~Isola, J.-Y. Zhu, T.~Zhou, and A.~A. Efros, ``Image-to-image translation with conditional adversarial networks,'' in {\em Proceedings of the IEEE conference on computer vision and pattern recognition (CVPR)}, 2017.

\bibitem{chollet2019measure}
F.~Chollet, ``On the measure of intelligence,'' {\em arXiv preprint arXiv:1911.01547}, 2019.

\bibitem{shanahan2020explicitly}
M.~Shanahan, K.~Nikiforou, A.~Creswell, C.~Kaplanis, D.~Barrett, and M.~Garnelo, ``An explicitly relational neural network architecture,'' in {\em International Conference on Machine Learning (ICML)}, PMLR, 2020.

\bibitem{raven}
C.~Zhang, F.~Gao, B.~Jia, Y.~Zhu, and S.-C. Zhu, ``Raven: A dataset for relational and analogical visual reasoning,'' in {\em Proceedings of the IEEE/CVF conference on computer vision and pattern recognition (CVPR)}, pp.~5317--5327, 2019.

\bibitem{iraven}
S.~Hu, Y.~Ma, X.~Liu, Y.~Wei, and S.~Bai, ``Stratified rule-aware network for abstract visual reasoning,'' in {\em Proceedings of the AAAI Conference on Artificial Intelligence (AAAI)}, vol.~35, pp.~1567--1574, 2021.

\bibitem{susskind2021neuro}
Z.~Susskind, B.~Arden, L.~K. John, P.~Stockton, and E.~B. John, ``Neuro-symbolic ai: An emerging class of ai workloads and their characterization,'' {\em arXiv preprint arXiv:2109.06133}, 2021.

\bibitem{kleyko2021vector}
D.~Kleyko {\em et~al.}, ``{Vector Symbolic Architectures as a Computing Framework for Emerging Hardware},'' {\em Proceedings of the IEEE}, vol.~110, no.~10, pp.~1538--1571, 2022.

\bibitem{frady2020resonator}
E.~P. Frady {\em et~al.}, ``{Resonator Networks, 1: An Efficient Solution for Factoring High-Dimensional, Distributed Representations of Data Structures},'' {\em Neural Computation}, vol.~32, no.~12, pp.~2311--2331, 2020.

\bibitem{wan2024h3dfact}
Z.~Wan, C.-K. Liu, M.~Ibrahim, H.~Yang, S.~Spetalnick, T.~Krishna, and A.~Raychowdhury, ``H3dfact: Heterogeneous 3d integrated cim for factorization with holographic perceptual representations,'' in {\em 2024 Design, Automation \& Test in Europe Conference \& Exhibition (DATE)}, pp.~1--6, IEEE, 2024.

\bibitem{xu2024ferex}
Z.~Xu, C.-K. Liu, C.~Li, R.~Mao, J.~Yang, T.~K{\"a}mpfe, M.~Imani, C.~Li, C.~Zhuo, and X.~Yin, ``Ferex: A reconfigurable design of multi-bit ferroelectric compute-in-memory for nearest neighbor search,'' in {\em Design, Automation \& Test in Europe Conference \& Exhibition (DATE), 2024}, IEEE, 2024.

\bibitem{shou2023see}
S.~Shou, C.-K. Liu, S.~Yun, Z.~Wan, K.~Ni, M.~Imani, X.~S. Hu, J.~Yang, C.~Zhuo, and X.~Yin, ``See-mcam: Scalable multi-bit fefet content addressable memories for energy efficient associative search,'' in {\em 2023 IEEE/ACM International Conference on Computer Aided Design (ICCAD)}, pp.~1--9, IEEE, 2023.

\bibitem{kleyko2022vector}
D.~Kleyko, M.~Davies, E.~P. Frady, P.~Kanerva, S.~J. Kent, B.~A. Olshausen, E.~Osipov, J.~M. Rabaey, D.~A. Rachkovskij, A.~Rahimi, {\em et~al.}, ``Vector symbolic architectures as a computing framework for emerging hardware,'' {\em Proceedings of the IEEE}, vol.~110, no.~10, 2022.

\bibitem{kanerva2010we}
P.~Kanerva, ``{Prototypes and Mapping in Concept Space},'' in {\em Proceedings of AAAI Fall Symposium: Quantum Informatics for Cognitive Social and Semantic Processes}, pp.~2--6, 2010.

\bibitem{kleyko2021cellular}
D.~Kleyko, E.~P. Frady, and F.~T. Sommer, ``Cellular automata can reduce memory requirements of collective-state computing,'' {\em IEEE Transactions on Neural Networks and Learning Systems}, vol.~33, no.~6, 2021.

\bibitem{datta2019programmable}
S.~Datta {\em et~al.}, ``{A Programmable Hyper-Dimensional Processor Architecture for Human-Centric IoT},'' {\em IEEE JETCAS}, vol.~9, no.~3, 2019.

\bibitem{menon2023shared}
A.~Menon {\em et~al.}, ``{Shared Control of Assistive Robots through User-intent Prediction and Hyperdimensional Recall of Reactive Behavior},'' in {\em Proc. IEEE ICRA}, pp.~12638--12644, 2023.

\bibitem{booch2021thinking}
G.~Booch, F.~Fabiano, L.~Horesh, K.~Kate, J.~Lenchner, N.~Linck, A.~Loreggia, K.~Murgesan, N.~Mattei, F.~Rossi, {\em et~al.}, ``Thinking fast and slow in ai,'' in {\em Proceedings of the AAAI Conference on Artificial Intelligence (AAAI)}, vol.~35, pp.~15042--15046, 2021.

\bibitem{kambhampati2024llms}
S.~Kambhampati, K.~Valmeekam, L.~Guan, K.~Stechly, M.~Verma, S.~Bhambri, L.~Saldyt, and A.~Murthy, ``Llms can't plan, but can help planning in llm-modulo frameworks,'' {\em arXiv preprint arXiv:2402.01817}, 2024.

\end{thebibliography}

\vspace{-0.5in}
% Zishen
\begin{IEEEbiography}[{\includegraphics[width=1in,height=1.25in,clip,keepaspectratio]{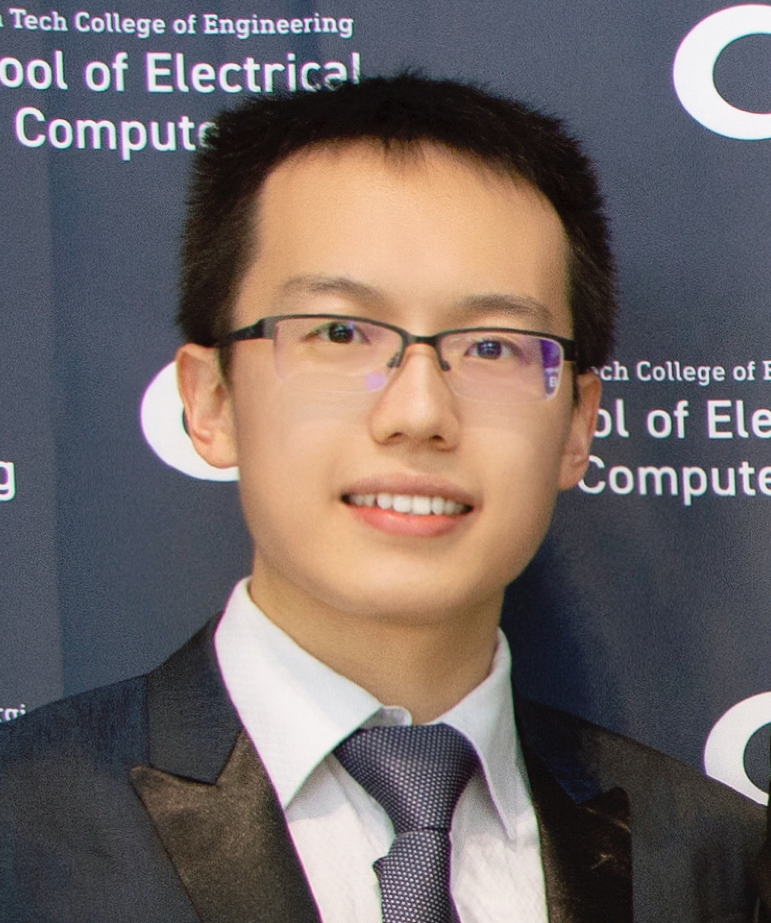}}]{Zishen Wan} (Student Member, IEEE) received the B.E. degree in electrical engineering and automation from Harbin Institute of Technology, Harbin, China, in 2018, and the M.S. degree in electrical engineering from Harvard University, Cambridge, MA, USA, in 2020. Currently, he is  pursuing Ph.D. degree with the School of Electrical and Computer Engineering, Georgia Institute of Technology, Atlanta, GA, USA.

His research interests include computer architecture, VLSI, and embedded systems, with a focus on designing efficient and reliable hardware and systems for autonomous machines and cognitive intelligence.
\end{IEEEbiography}

\vspace{-0.5in}

% Che-Kai
\begin{IEEEbiography}[{\includegraphics[width=1in,height=1.25in,clip,keepaspectratio]{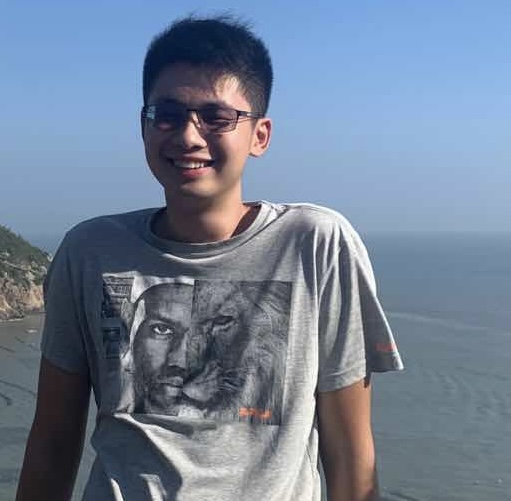}}]{Che-Kai Liu} (Student Member, IEEE) received the B.Eng. degree in Electronic Engineering from Zhejiang University, Hangzhou, China, in 2023. He is currently pursuing Ph.D. degree at the Integrated Circuits and Systems Research Lab (ICSRL), Georgia Institute of Technology, USA, under the supervision of Dr. Arijit Raychowdhury. He currently holds a research intern position at Corporate Research, TSMC, San Jose, USA. 

His research interest includes SoC prototype for next-generation AI applications.
\end{IEEEbiography}

\vspace{-0.5in}

% Hanchen
\begin{IEEEbiography}[{\includegraphics[width=1in,height=1.25in,clip,keepaspectratio]{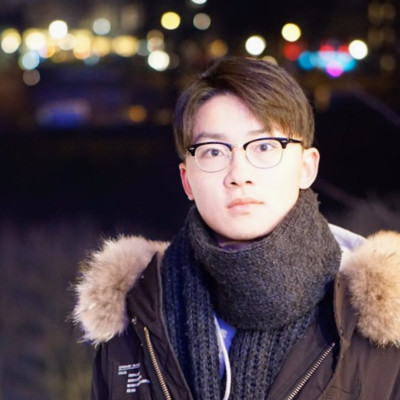}}]{Hanchen Yang} (Student Member, IEEE) received the B.S. degree in applied physics from Beijing University of Posts and Telecommunications, Beijing, China, in 2018, and the M.S. degree in electrical and computer engineering from Carnegie Mellon University, Pittsburgh, PA, USA, in 2020. Currently, he is pursuing Ph.D. degree with the School of Electrical and Computer Engineering, Georgia Institute of Technology, Atlanta, GA, USA.

His research interests include computer architecture, neural-symbolic AI, and hardware-software co-design for novel ML applications.
\end{IEEEbiography}

\vspace{-0.5in}

% Ritik
\begin{IEEEbiography}[{\includegraphics[width=1in,height=1.25in,clip,keepaspectratio]{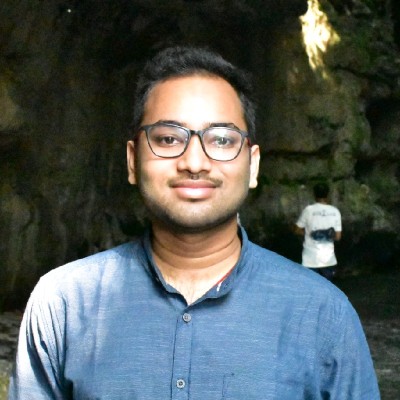}}]{Ritik Raj} (Student Member, IEEE) received the B.Tech. degree in electronics and communication engineering from Indian Institute of Technology, Roorkee, India, in 2023. Currently, he is  pursuing Ph.D. degree with the School of Electrical and Computer Engineering, Georgia Institute of Technology, Atlanta, GA, USA.

His research interests include computer architecture, specifically AI and domain-specific accelerators, FPGA, and architecture simulator design.
\end{IEEEbiography}

\vspace{-0.5in}

% Chaojian
\begin{IEEEbiography}[{\includegraphics[width=1in,height=1.25in,clip,keepaspectratio]{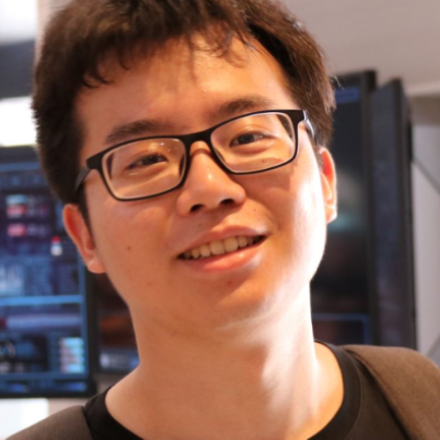}}]{Chaojian Li} (Student Member, IEEE) received the B.E. degree in precision instrument from Tsinghua University, Beijing, China, in 2019. Currently, he is  pursuing Ph.D. degree with the School of Computer Science, Georgia Institute of Technology, Atlanta, GA, USA.

His research interests include deep learning and computer architecture, with a focus on 3D reconstruction and rendering in an algorithm-hardware co-design approach and deep learning on edge devices.
\end{IEEEbiography}

\vspace{-0.5in}

% Haorao
\begin{IEEEbiography}[{\includegraphics[width=1in,height=1.25in,clip,keepaspectratio]{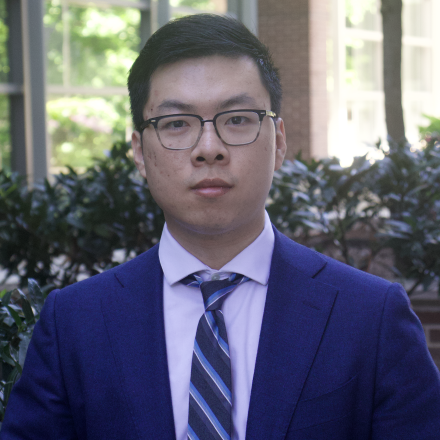}}]{Haoran You} (Student Member, IEEE) received the B.E. degree in electronic information and communication from Huazhong University of Science and Technology, Wuhan, China, in 2019. Currently, he is pursuing Ph.D. degree with the School of Computer Science, Georgia Institute of Technology, Atlanta, GA, USA.

His research interests are efficient and automated ML/AI systems through algorithm-hardware co-design.
\end{IEEEbiography}

\vspace{-0.5in}

% Yonggan
\begin{IEEEbiography}[{\includegraphics[width=1in,height=1.25in,clip,keepaspectratio]{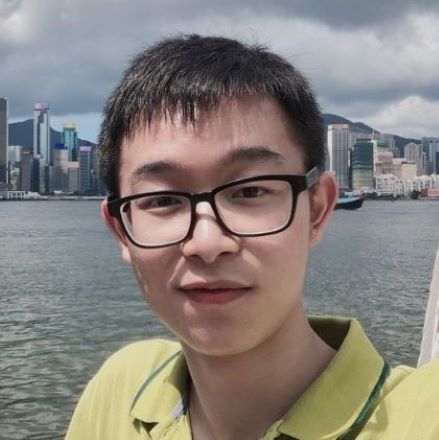}}]{Yonggan Fu} (Student Member, IEEE) received the B.E. degree in applied physics and computer science from University of Science and Technology of China, Hefei, China, in 2019. Currently, he is pursuing Ph.D. degree with the School of Computer Science, Georgia Institute of Technology, Atlanta, GA, USA.

His research interests are developing efficient and robust AI algorithms and co-designing the corresponding hardware accelerators towards a triple-win in accuracy, efficiency, and robustness.
\end{IEEEbiography}

\vspace{-0.5in}

% Cheng
\begin{IEEEbiography}[{\includegraphics[width=1in,height=1.25in,clip,keepaspectratio]{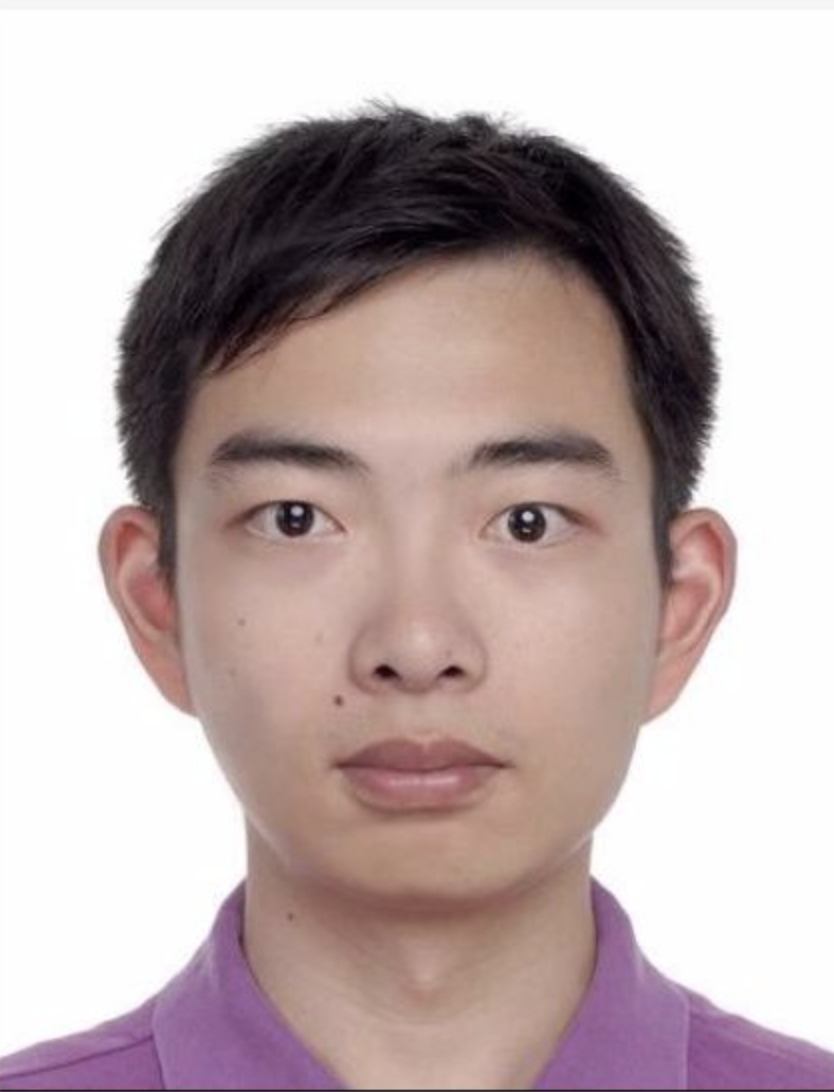}}]{Cheng Wan} (Student Member, IEEE) received the B.E. degree in computer science at Shanghai Jiao Tong University, Shanghai, China, in 2018. Currently, he is pursuing Ph.D. degree with the School of Computer Science, Georgia Institute of Technology, Atlanta, GA, USA.

His research interests include algorithm-system co-design for machine learning systems, with a special focus on distributed training.
\end{IEEEbiography}

\vspace{-0.5in}

% Sixu
\begin{IEEEbiography}[{\includegraphics[width=1in,height=1.25in,clip,keepaspectratio]{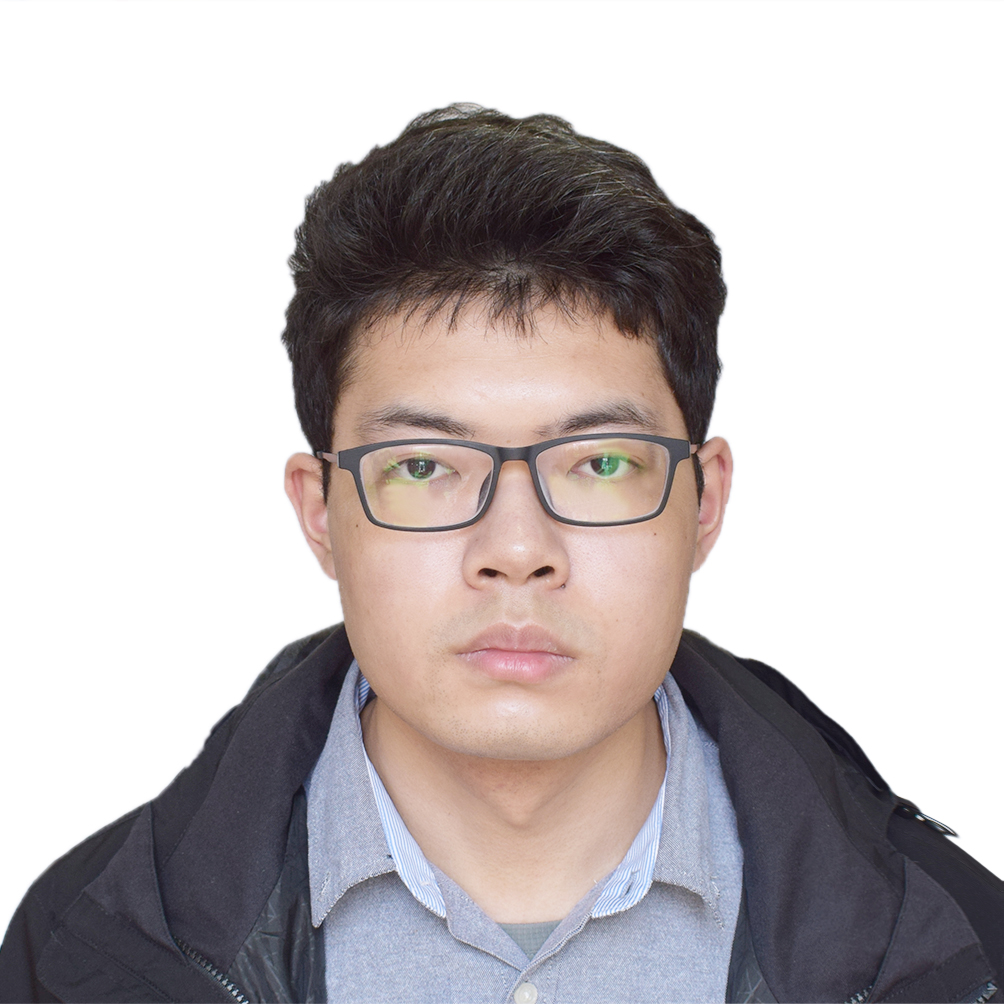}}]{Sixu Li} (Student Member, IEEE) received the B.E. degree in communication engineering from the University of Electronic Science and Technology of China, Chengdu, China, in 2021. Currently, he is pursuing Ph.D. degree with the School of Computer Science, Georgia Institute of Technology, Atlanta, GA, USA.

His research interests include computer architecture, digital circuits and systems, with a focus on accelerator design for neural rendering.
\end{IEEEbiography}

\vspace{-0.5in}

% Youbin
\begin{IEEEbiography}[{\includegraphics[width=1in,height=1.25in,clip,keepaspectratio]{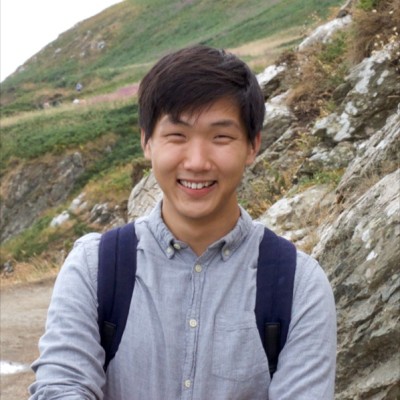}}]{Youbin Kim} (Student Member, IEEE) received the B.A. degree in physics from Harvard University, Cambridge, MA, USA, in 2018. Currently, he is pursuing Ph.D. degree with the School of Electrical Engineering, University of California, Berkeley, Berkeley, CA, USA.

His research interests include digital integrated circuits, computer architecture, and the design of hardware accelerators for high-dimensional computing and machine learning algorithms.
\end{IEEEbiography}

\vspace{-0.5in}

% Anand
\begin{IEEEbiography}[{\includegraphics[width=1in,height=1.25in,clip,keepaspectratio]{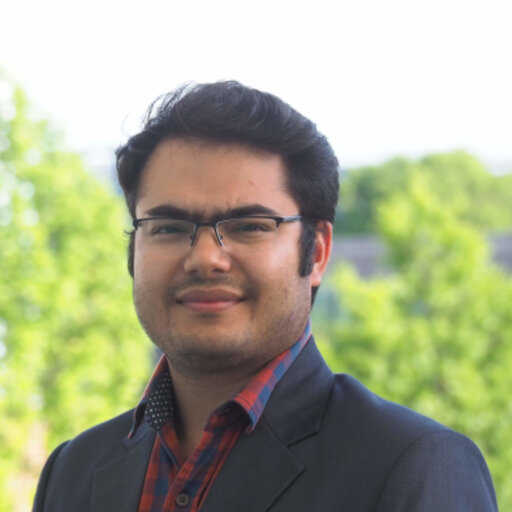}}]{Ananda Samajdar} (Member, IEEE) received the Ph.D. degree in electrical and computer engineering, Georgia Institute of Technology, Atlanta, GA, USA, in 2022. Currently, he is the research staff member at IBM T.J. Watson Research Center, Yorktown Heights, NY, USA.

His research interests include computer archtitecture, VLSI design, computer systems design, machine learning algorithm development.
\end{IEEEbiography}

\vspace{-0.5in}

% Celine
% \begin{IEEEbiography}[{\includegraphics[width=1in,height=1.25in,clip,keepaspectratio]{Figs/author/Celine.jpg}}]{Yingyan (Celine) Lin} (Member, IEEE) received the Ph.D. degree in electrical and computer engineering from the University of Illinois at Urbana-Champaign, Champaign, IL, in 2017. Currently, she is the associate professor at the School of Computer Science, Georgia Institute of Technology, Atlanta, GA, USA.

% Dr. Lin's research interests include efficient machine learning systems via cross-layer innovations, aiming to develop efficient algorithms, accelerators, and automated tools for enabling ubiquitous on-device intelligence and promoting green AI. She was a recipient of the NSF CAREER Award, the IBM Faculty Award, and the Meta Faculty Research Award and recently received the ACM SIGDA Outstanding Young Faculty Award. She served as the Program Co-Chair for the 32nd IEEE International Conference on Application-Specific Systems, Architectures and Processors (ASAP 2021). She is currently a Track Chair for the ACM/EDAC/IEEE Design Automation Conference (DAC) and an Associate Editor of IEEE TRANSACTIONS ON CIRCUITS AND SYSTEMS— II: EXPRESS BRIEFS.
\begin{IEEEbiography}[{\includegraphics[width=1in,height=1.25in,clip,keepaspectratio]{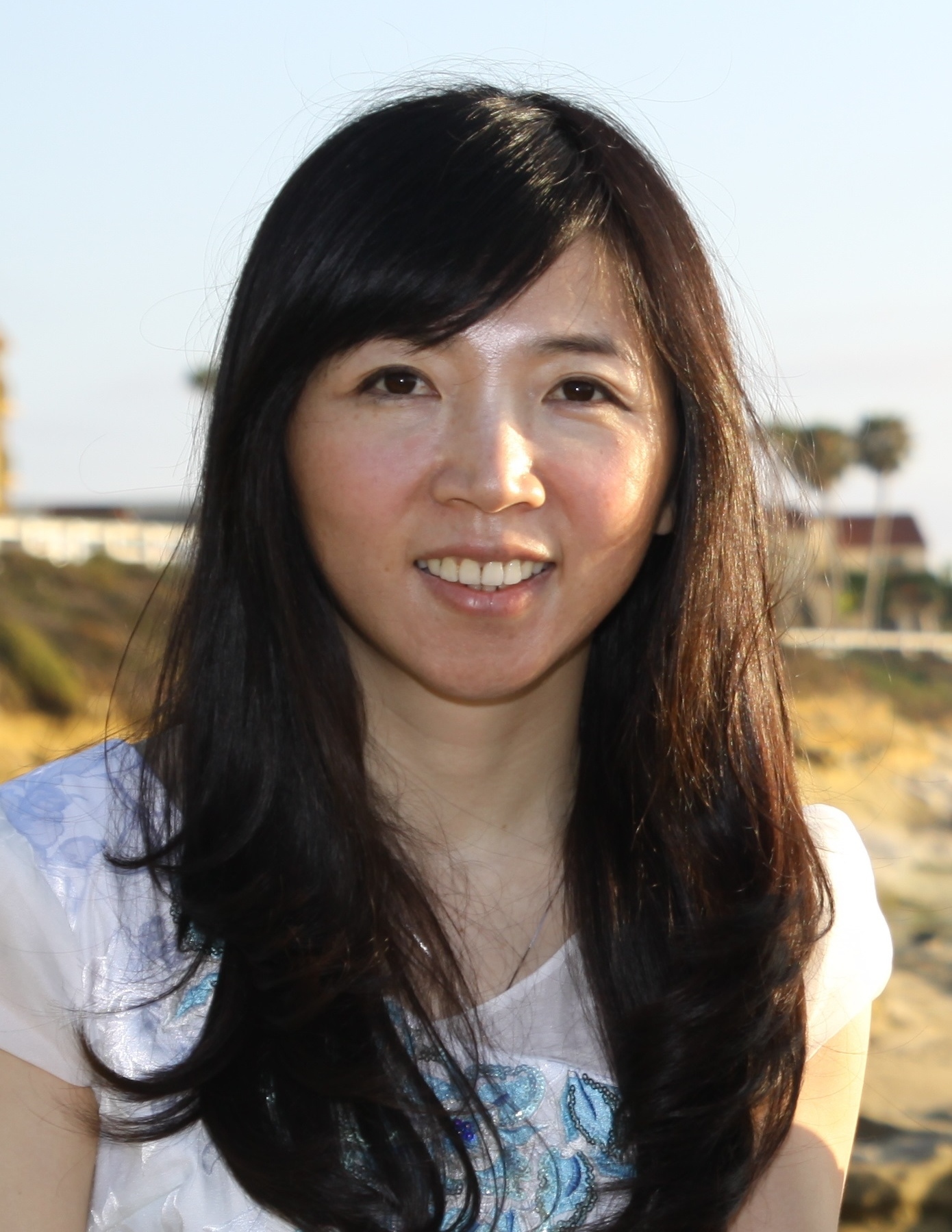}}]{Yingyan (Celine) Lin} (Member, IEEE) received her Ph.D. degree in Electrical and Computer Engineering from the University of Illinois at Urbana-Champaign, Champaign, IL, in 2017. Currently, she is an Associate Professor at the School of Computer Science and also serves as the Co-Director of the Center for Advancing Responsible Computing (CARE) at the Georgia Institute of Technology, Atlanta, GA, USA.

Dr. Lin's research focuses on developing efficient machine learning solutions through cross-layer innovations from efficient AI algorithms to AI accelerator and chip design in order to enable ubiquitous on-device intelligence and promote green AI. Her work has won first place in both the University Demonstration at DAC 2022 and the ACM/IEEE TinyML Design Contest at ICCAD 2022, and was selected as an IEEE Micro Top Pick of 2023. She has received the NSF CAREER Award, the IBM Faculty Award, the Meta Faculty Research Award, the ACM SIGDA Outstanding Young Faculty Award, and recently the SRC Young Faculty Award. She is currently serving as the Program Co-Chair for the eighth annual Machine Learning and Systems conference (MLSys 2025) and on the Technical Program Committees of several first-tier conferences, such as DAC and MICRO.
\end{IEEEbiography}

\vspace{-0.5in}

% Mohamed
\begin{IEEEbiography}[{\includegraphics[width=1in,height=1.25in,clip,keepaspectratio]{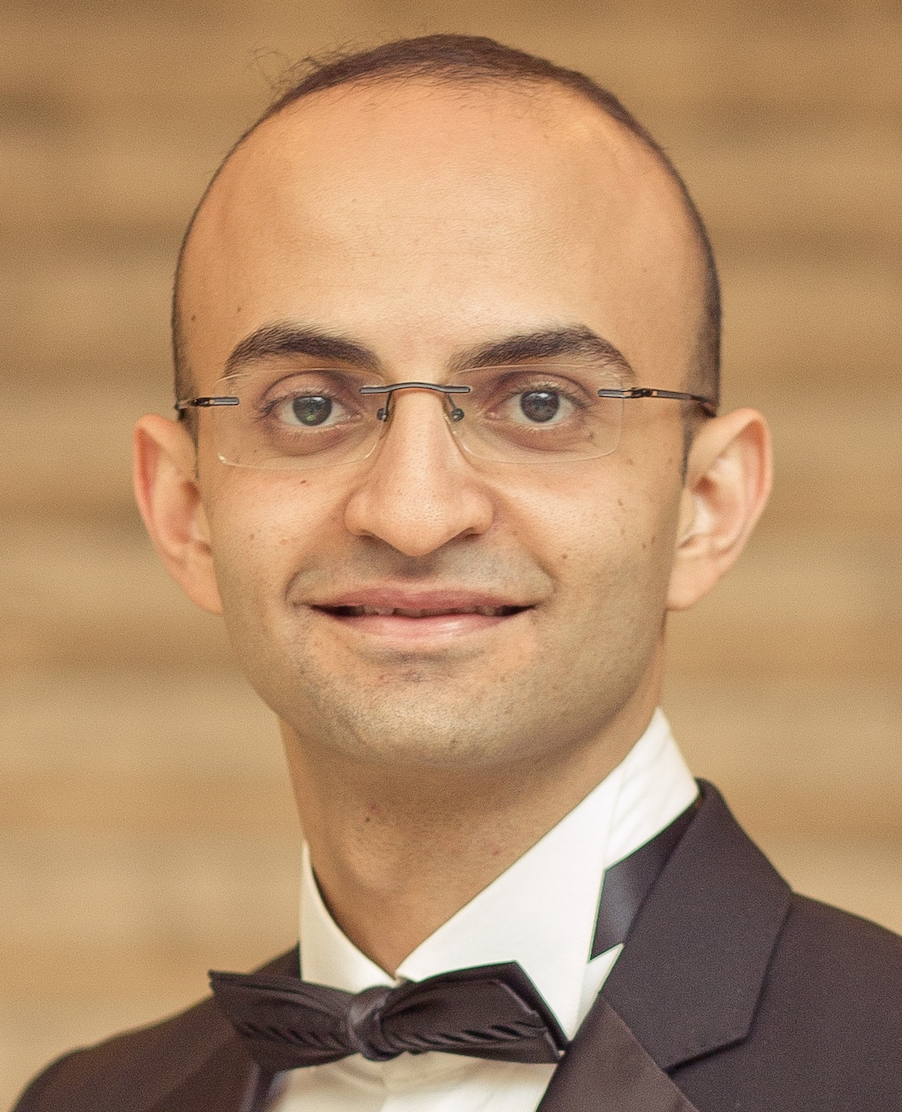}}]{Mohamed Ibrahim} (Member, IEEE) received the Ph.D. degree in electrical and computer engineering from Duke University, Durham, NC, in 2018. He was a postdoctoral researcher at the University of California at Berkeley, Berkeley, CA, USA. Currently, he is a senior research engineer and research faculty member with the School of Electrical and Computer Engineering, Georgia Institute of Technology, Atlanta, GA, USA.

His research is broadly at the intersection of energy-efficient AI, brain-inspired computing, and human-centered interactive systems.
\end{IEEEbiography}

\vspace{-0.4in}

% Jan
\begin{IEEEbiography}[{\includegraphics[width=1in,height=1.25in,clip,keepaspectratio]{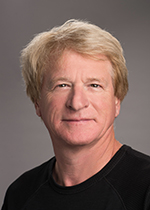}}]{Jan M. Rabaey} (Life Fellow, IEEE) is a professor with Graduate School, Electrical Engineering and Computer Science Department, the University of California at Berkeley, Berkeley, CA, USA, after being the holder of the Donald O. Pederson Distinguished Professorship at the same institute for over 30 years. He is a founding director of the Berkeley Wireless Research Center (BWRC) and the Berkeley Ubiquitous SwarmLab, and has served as the Electrical Engineering Division Chair at Berkeley twice. In 2019, he also became the CTO of the System-Technology Co-Optimization (STCO) Division of IMEC, Belgium.

Dr. Rabaey has made high-impact contributions to a number of fields, including low power integrated circuits, advanced wireless systems, mobile devices, sensor networks, and ubiquitous computing.  Some of the systems he helped envision include the infoPad (a forerunner of the iPad), PicoNets and PicoRadios (IoT avant-la-lettre), the Swarm (IoT on steroids), Brain-Machine interfaces and the Human Intranet. His current interests include the conception of the next-generation distributed systems, as well as the exploration of the interaction between the cyber and the biological worlds. He is the primary author of the influential \textit{Digital Integrated Circuits: A Design Perspective} textbook that has served to educate hundreds of thousands of students all over the world. He is the recipient of numerous awards, is a Life Fellow of the IEEE, and has been involved in a broad variety of start-up ventures.

\end{IEEEbiography}

\vspace{-0.6in}

% Tushar
\begin{IEEEbiography}[{\includegraphics[width=1in,height=1.25in,clip,keepaspectratio]{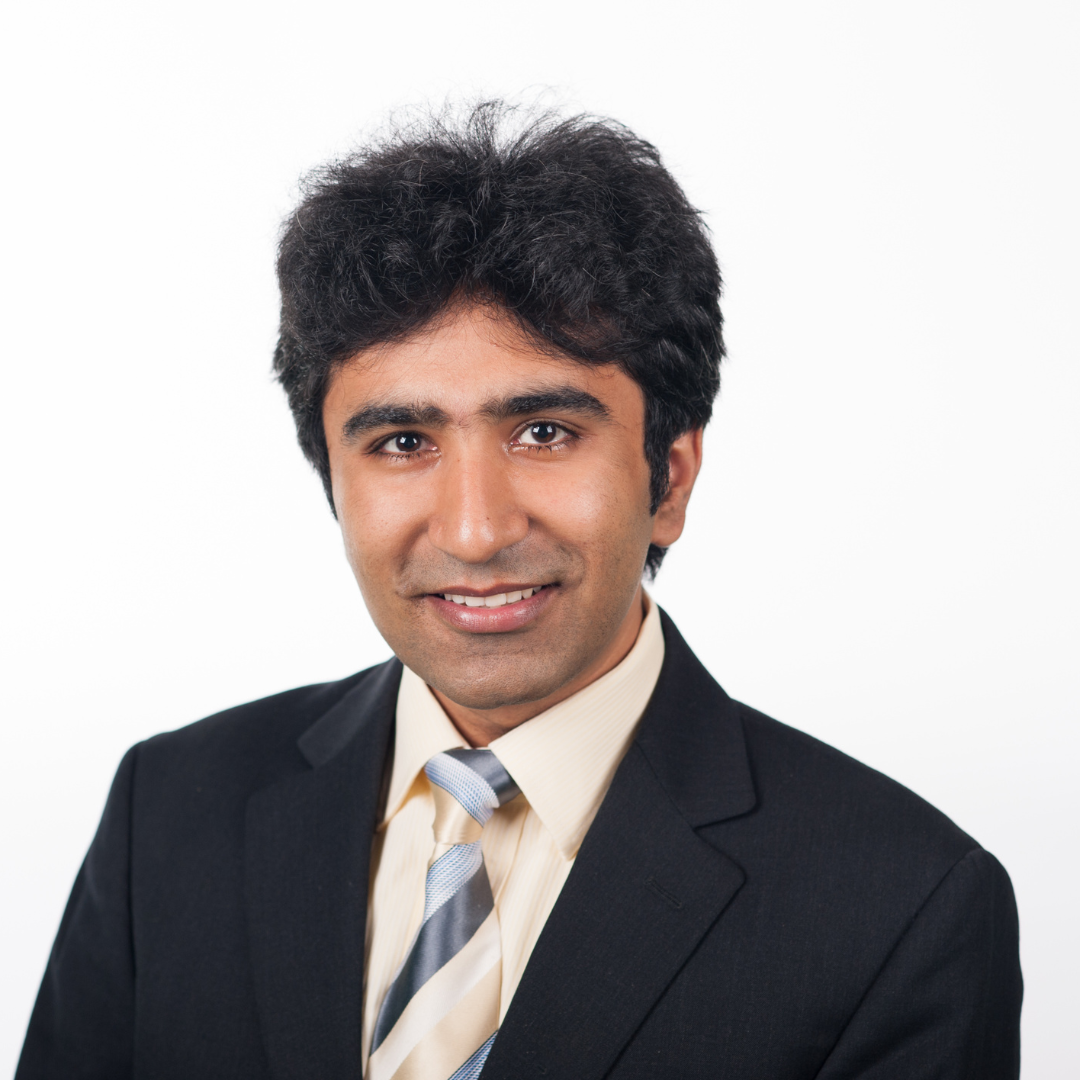}}]{Tushar Krishna} (Senior Member, IEEE) received the Ph.D. degree in electrical engineering and computer science from Massachusetts Institute of Technology, Cambridge, MA, USA, in 2014. Currently, he is an Associate Professor with the School of Electrical and Computer Engineering, Georgia Institute of Technology, Atlanta, GA, USA.

Dr. Krishna’s research spans computer architecture, interconnection networks, networks-on-chip, and AI/ML accelerator systems – with a focus on optimizing data movement in modern computing platforms. His research is funded via multiple awards from NSF, DARPA, IARPA, SRC (including JUMP2.0), Department of Energy, Intel, Google, Meta/Facebook, Qualcomm and TSMC. Three of his papers have been selected for IEEE Micro’s Top Picks from Computer Architecture, one more received an honorable mention, and four have won best paper awards. Dr. Krishna was inducted into the HPCA Hall of Fame in 2022. He has been honored by the Roger P. Webb Outstanding Junior Faculty Award in 2021, the Richard M. Bass/Eta Kappa Nu Outstanding Junior Teacher Award in 2023, and the Roger P. Webb Outstanding Mid-career Faculty Award in 2024.
\end{IEEEbiography}

\vspace{-0.4in}

% Arijit
\begin{IEEEbiography}[{\includegraphics[width=1in,height=1.25in,clip,keepaspectratio]{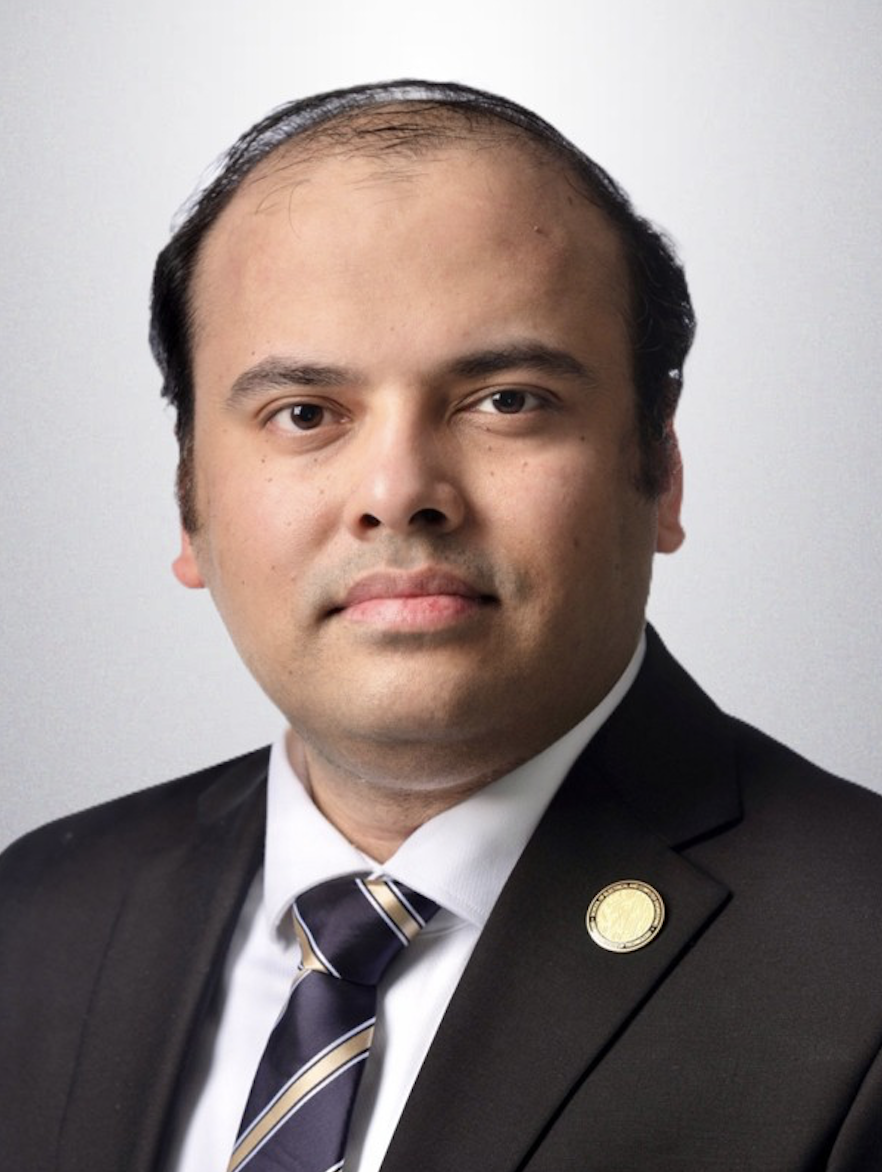}}]{Arijit Raychowdhury} (Fellow, IEEE) received the Ph.D. degree in electrical and computer engineering from Purdue University, West Lafayette, IN, USA, in 2007. He is currently the Steve W Chaddick Chair and a Professor with the School of Electrical and Computer Engineering, Georgia Institute of Technology, Atlanta, GA, USA. He is also the Director of the Center for the Co-Design of Cognitive Systems (CoCoSys), a Joint University Microelectronics Program 2.0. His research interests include low-power digital and mixed-signal circuit design, signal processors, and exploring interactions of circuits with device technologies. 

Dr. Raychowdhury is currently a Distinguished Lecturer of the IEEE Solid State Circuits Society (SSCS). He serves on the Technical Program Committee of key circuits and design conferences, including ISSCC, VLSI Symposium, DAC, and CICC. He is the winner of several awards, including the SRC Technical Excellence Award in 2021, the Qualcomm Faculty Award in 2021 and 2020, the IEEE/ACM Innovator under 40 Award, and the NSF CISE Research Initiation Initiative Award (CRII) in 2015.
\end{IEEEbiography}


% \end{IEEEbiography}


%\newpage



%\vfill


\end{document}